\documentclass[modern]{aastex62}
\usepackage[normalem]{ulem}
\usepackage{makecell}
\usepackage{graphics}
\usepackage{amsmath}
\usepackage{soul}
\newcommand{\halpha}{{\rm H}\alpha}
\newcommand{\lhalbol}{L_{{\rm H}\alpha}/L_{\rm bol}}
\newcommand{\lxraylbol}{L_{{\rm Xray}}/L_{\rm bol}}
\newcommand{\gbp}{{G_\mathrm{BP}}}
\newcommand{\grp}{{G_\mathrm{RP}}}

\received{December 14 2018}
\accepted{April 5 2019}
\submitjournal{The Astronomical Journal}

\begin{document}
\title{Exploring the age dependent properties of M and L dwarfs using \textit{Gaia} and SDSS}
\shorttitle{M and L dwarfs in \textit{Gaia} and SDSS}

\author[0000-0003-2102-3159]{Rocio Kiman}
\affil{Department of Physics, Graduate Center, City University of New York, 365 5th Ave, New York, NY 10016, USA}
\affil{Department of Astrophysics, American Museum of Natural History, Central Park West at 79th St, New York, NY 10024, USA}
\email{rociokiman@gmail.com}

\author[0000-0002-7224-7702]{Sarah J. Schmidt}
\affil{Leibniz-Institute for Astrophysics Potsdam (AIP), An der Sternwarte 16, 14482, Potsdam, Germany}

\author[0000-0003-4540-5661]{Ruth Angus}
\affil{Department of Astrophysics, American Museum of Natural History, Central Park West at 79th St, New York, NY 10024, USA}
\affil{Center for Computational Astrophysics, Flatiron Institute, 162 5th Avenue, New York, NY 10010 USA}
\affil{Department of Astronomy, Columbia University, 116th St \& Broadway, New York, NY 10027, USA}

\author[0000-0002-1821-0650]{Kelle L. Cruz}
\affil{Department of Physics, Graduate Center, City University of New York, 365 5th Ave, New York, NY 10016, USA}
\affil{Department of Astrophysics, American Museum of Natural History, Central Park West at 79th St, New York, NY 10024, USA}
\affil{Center for Computational Astrophysics, Flatiron Institute, 162 5th Avenue, New York, NY 10010 USA}
\affil{Hunter College, City University of New York, 695 Park Ave, New York, NY 10065, USA}

\author[0000-0001-6251-0573]{Jacqueline K. Faherty}
\affil{Department of Astrophysics, American Museum of Natural History, Central Park West at 79th St, New York, NY 10024, USA}
\author[0000-0002-3252-5886]{Emily Rice}
\affil{Department of Physics, Graduate Center, City University of New York, 365 5th Ave, New York, NY 10016, USA}
\affil{Department of Astrophysics, American Museum of Natural History, Central Park West at 79th St, New York, NY 10024, USA}
\affil{College of Staten Island, City University of New York, 2800 Victory Blvd, Staten Island, NY 10314, USA}

\begin{abstract} 
We present a sample of $74,216$ M and L dwarfs\footnote{Find the sample here: https://zenodo.org/record/2636692\#.XK9\_1etKjVp} constructed from two existing catalogs of cool dwarfs spectroscopically identified in the Sloan Digital Sky Survey (SDSS). We  cross-matched the SDSS catalog with \textit{Gaia} DR2 to obtain parallaxes and proper motions and modified the quality cuts suggested by the Gaia Collaboration to make them suitable for late-M and L dwarfs. We also provide relations between \textit{Gaia} colors and absolute magnitudes with spectral type and conclude that $(G-\grp)$ has the tightest relation to spectral type for M and L dwarfs. In addition, we study magnetic activity as a function of position on the color--magnitude diagram, finding that $\halpha$ magnetically active stars have, on average, redder colors and/or brighter magnitudes than inactive stars. This effect cannot be explained by youth alone and might indicate that active stars are magnetically inflated, binaries and/or high metallicity. Moreover, we find that vertical velocity and vertical action dispersion are correlated with $\halpha$ emission, confirming that these two parameters are age indicators. We also find that stars below the main sequence have high tangential velocity which is consistent with a low metallicity and old population of stars that belong to the halo or thick disk. 
\end{abstract}


\section{Introduction}
\label{sec:intro}

The Milky Way Galaxy is dominated in number by low mass stars occupying the M and L spectral types \citep[e.g.][]{Gould1996,Bochanski2010}. M dwarfs have a wide range of ages in the Milky Way Galaxy since they have main-sequence lifetimes longer than the current age of the Universe \citep[e.g.][]{Fagotto1994,Laughlin1997}. This makes them an ideal population for studies of the structure, dynamics, and evolution of the stellar thin disk. Ages of field solar-type stars are typically obtained by three methods \citep{Soderblom2010}: (1) empirical methods such as activity--age relations \citep{Mamajek2008} and rotation period--age relations, called gyrochronology \citep{Skumanich1972,Barnes2007,Angus2015,Meibom2015,VanSaders2016}; (2) model-dependent methods such as isochrone fitting \citep{Edvardsson1993} and asteroseismology \citep{Chaplin2014}; and (3) statistical methods such as kinematic age dating \citep{Wielen1977}. 
Despite the availability of multiple methods, assigning accurate ages to M and L dwarfs in the field remains challenging. 
Due to their long main-sequence lifetimes, age-related parameters change slowly with time. 
Asteroseismological methods cannot yet be applied to M dwarfs as their acoustic oscillations have extremely small amplitudes \citep{Rodriguez2016}, and their isochronal stellar evolution models are not accurate, in part because of the difficulty associated with modeling fully convective interiors \citep{Baraffe2015}. Empirical and statistical methods are the best option to obtain ages for field M and L dwarfs. 

Solar-type stars have a radiative core and a convective envelope that do not rotate as a rigid body. 
It is generally thought that as a consequence of this differential rotation, a dynamo is generated at the interface of the two zones,
which is responsible for the magnetic activity of the star \citep{Parker1955}. 
When the magnetic field threads through the surface, it heats the chromosphere and the corona, generating collisionally induced atomic emission (including the $\halpha$ emission line) and X-ray emission, respectively. 
As a consequence, $\halpha$ and X-ray emission are measurable evidence of surface magnetism that can be used as magnetic activity indicators.
The magnetic field is also partly responsible for the stellar magnetic wind, which dissipates angular momentum, slowing the rotation (and thus the differential rotation) of the star. Due to this process, rotation, magnetic activity and age are tightly related for solar-type stars \citep[e.g.][]{Skumanich1972,Barry1988,Soderblom1991,Mamajek2008}. Stars with masses $<0.35$~M$_\odot$ (spectral type $\sim$M3) are fully convective \citep{Chabrier1997}, so there is no interface with a radiative zone to produce a solar-type dynamo. 
Even though the mechanism to generate magnetic fields in fully convective stars is not yet understood, a strong correlation between rotation and magnetic activity is found for fully convective M dwarfs  \citep[e.g.][]{Delfosse1998,Mohanty2003,Reiners2012,West2015,Newton2017}. Furthermore, several studies have extended the idea that magnetic activity decreases with age for late-M dwarfs \citep{Fleming1995,Eggen1990,West2006,West2008a,RIEDEL2016}. This indicates that there is an empirical relation between age, rotation and magnetic activity for M dwarfs that may extend to L dwarfs as well.

As they orbit the center of the Galaxy, stars gravitationally interact with giant molecular clouds and other passing stars, receiving a kinematic kick that alters their orbits. The increased eccentricity and inclination of the altered orbits causes the stars to separate from the plane of the Galaxy as they age. This effect is generally quantified by the age-velocity relation \citep[AVR;][]{Wielen1977,Hanninen2002}, which indicates the velocity dispersion of a population of stars with a similar age, goes as the square root of its age ($\sigma = t^{0.5}$). This relationship is particularly strong when examining the correlations between Galactic height or vertical velocity and age \citep[e.g.][]{West2006,West2008a,Nordstrom2004,Aumer2016,Yu2018}. This statistical method was used by several works to obtain kinematic ages of population of stars \citep[e.g.][]{Schmidt2007,ZapateroOsorio2007,Faherty2009,Reiners2009}. Thanks to large spectrophotometric surveys such as the Sloan Digital Sky Survey \citep[SDSS;][]{York2000} and astrometric surveys such as \textit{Gaia} DR2 \citep{GaiaCollaboration2016,Collaboration2018}, better results can be expected from statistical methods.



Our ultimate goal is to infer the ages of M and L dwarfs by combining different age indicators such as fractional $\halpha$ luminosity and vertical action dispersion. 
We began this process by compiling a sample of tens of thousands of spectroscopically identified M and L dwarfs, including colors, activity measurements, and kinematics with sufficient precision to use vertical action dispersion as an age indicator. In this paper, we introduce the MLSDSS-GaiaDR2 sample which includes H$\alpha$ equivalent widths, spectral types for M and L dwarfs, and radial velocities from two catalogs compiled from SDSS: the spectroscopic M~dwarf catalog \citep{West2011} and the the ``BUD'' catalog of \citet[][in prep]{Schmidt2015,schmidt2019}; as well as their vertical velocities and actions, calculated from \textit{Gaia} DR2 proper motions, parallaxes and positions. 

This paper is laid out as follows.
In section \ref{sec:MLSDSS}, we describe the assembly of our M and L dwarf sample, including the process of cross-matching and combining data from different surveys, and the quality cuts we applied that
remove incorrect matches and low quality data.
In section \ref{sec:mlgaiadr2_colors_absmag}, we fit relations to the \textit{Gaia} colors/absolute magnitudes and spectral types of the M and L dwarfs in our catalog.
In section \ref{sec:mlinsdss2mass}, we fit relations between the Two Micron All Sky Survey \citep[2MASS;][]{Skrutskie2006} and SDSS absolute magnitudes and spectral types of stars in our sample. In section \ref{sec:age}, we briefly explore the relation between fractional $\halpha$ luminosity and tangential velocity and the position of the star in the \textit{Gaia} color--magnitude diagram. We find that magnetically active stars have redder colors and/or brighter magnitudes than inactive stars and show that this effect cannot be explained only by youth and that radius inflation, metallicity and binarity could be the causes. We also use color--magnitude position and tangential velocity to identify a possibly old halo or thick disk population of M dwarfs. In this section we also discuss the relation between three age indicators in our catalog: H$\alpha$ luminosity, vertical velocity, and vertical action. Finally, in section \ref{sec:summary} we summarize the work and our conclusions.

\section{The MLSDSS-GaiaDR2 Sample}
\label{sec:MLSDSS}


In this paper, we present the MLSDSS-GaiaDR2 sample of M and L dwarfs including spectral types, H$\alpha$ measurements, survey photometry, and Galactic kinematics. The compilation of the MLSDSS-GaiaDR2 sample was accomplished in two parts: assembling the base sample, dubbed the ``MLSDSS'' sample, and then cross-matching it with the \textit{Gaia} Data Release 2 \citep[DR2;][]{Collaboration2018}. The MLSDSS sample is based on data from the Sloan Digital Sky Survey Data Releases 7, 10, and 12 \citep[DR7; DR10; DR12;][]{Abazajian2009, Gott2014,Alam2015} and the SDSS-III Baryon Oscillation Spectroscopic Survey \citep[BOSS;][]{Dawson2013, Eisenstein2011}. The construction of the MLSDSS sample is described in Section~\ref{subsec:basemlsdss} and the cross-match with \textit{Gaia}~DR2 is in Section~\ref{subsec:crossmatch}. 
In Section \ref{subsec:qualitycuts}, we describe the criteria we applied to the MLSDSS-GaiaDR2 sample to create a high-quality astrometric sample and the three resulting photometric subsamples.

\subsection{Assembling the MLSDSS Sample}
\label{subsec:basemlsdss}

The MLSDSS sample is the combination of two catalogs of low mass stars identified in SDSS: the DR7 spectroscopic M~dwarf catalog \citep{West2011} and the BOSS Ultracool dwarf ``BUD'' late-M and L dwarf catalog of \citet[][in prep]{Schmidt2015,schmidt2019}. 

The DR7 spectroscopic M dwarf catalog contains $70,841$ M0--M9 dwarfs from SDSS DR7; these stars comprise the bulk of the MLSDSS sample. \citet{West2011} selected sources using color cuts designed to include all M dwarfs ($r-z>0.42$ and $i-z>0.24$) and then combined a spectral template matching code with visual inspection to classify stars with spectral types M0 to M9 based on their red-optical SDSS low-resolution ($R1800-2200$) spectra. They also measured the $\halpha$ equivalent width ($\halpha$~EW) and fractional luminosity ($\lhalbol$) from the SDSS spectra and included the values with uncertainties for each star in their catalog. Finally, they give SDSS $ugriz$ photometry and Two Micron All Sky Survey \citep[2MASS;][]{Skrutskie2006} $JHK_s$ photometry for the M dwarfs.

The second component of the MLSDSS sample is the BUD catalog composed of $12,998$~M7--L8 dwarfs. It includes $9,623$~M7--M9 dwarfs from the DR7 M dwarf catalog and an additional $484$~L dwarfs from SDSS DR7 \citep{Schmidt2010}. The BUD catalog was also complemented with $2,891$~M7--L8 dwarfs selected as ancillary targets in the BOSS survey and released in SDSS~DR10 \citep{Schmidt2015} and DR12~\citep[][in prep]{schmidt2019}. The additional M7--M9 dwarfs were selected using the same color-cuts of the \citet{West2011} catalog ($r-z>0.42$ and $i-z>0.24$); the L dwarfs were selected with $(i-z) > 1.44$. \cite{Schmidt2015} assigned spectral types for the L~dwarfs and the additional M~dwarfs. They adopted the spectral type classification assigned by \cite{West2011} for the rest of the M0--M8 dwarfs but re-classified all of the M9 dwarfs \citep[][in prep.]{schmidt2019}. 
\cite{Schmidt2015} measured $\halpha$ equivalent widths and $\lhalbol$ for all objects in the BUD catalog.
They also re-queried SDSS and 2MASS and reported new $r$, $i$, $z$, $J$, $H$ and $K_s$ photometry. 
For the $9,623$ M7--M9 objects which are present in both the DR7 M dwarf and BUD catalogs, we adopted the photometry, $\halpha$ EW, and spectral type from BUD into the MLSDSS sample.

\begin{figure}[ht!]
\centering
\includegraphics[width=\linewidth]{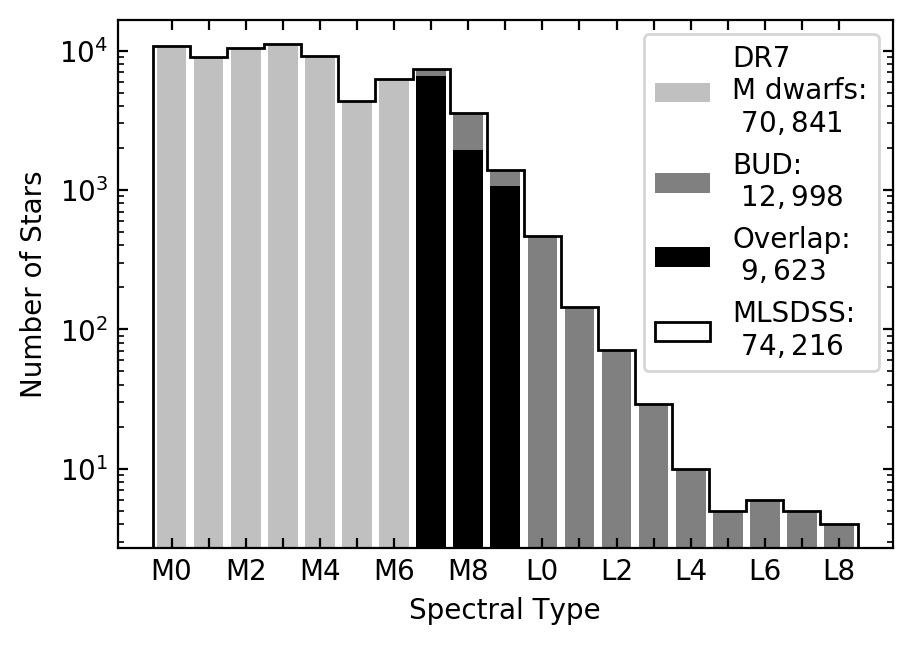}
\caption{Spectral type distribution of the MLSDSS sample (black outline). The objects from the SDSS DR7 M dwarf spectroscopic catalog are shown with light gray bars while dark gray bars indicate the objects from the BUD catalog. The overlapping objects from both catalogs are indicated by black bars.}
\label{fig:spt_MLSDSS}
\end{figure}

The spectral type distribution of the MLSDSS sample is shown in Figure~\ref{fig:spt_MLSDSS}.
The sample is not complete and reflects the SDSS target selection and sensitivities. The spectroscopic targeting of SDSS avoided some of the most common M3/M4 stars, and the sample is also incomplete at later spectral types and fainter magnitudes due to the capabilities of the telescope and instrument.

Both the SDSS DR7 M dwarf spectroscopic catalog and the BUD catalog contain data relevant to kinematics and activity: an activity field ($\texttt{ACTHA}$) and proper motion and radial velocity estimates.
We adopted these three values from the SDSS DR7 M dwarfs spectroscopic catalog for spectral types M0 to M6 and from the BUD catalog for spectral types M7 and later.

The $\texttt{ACTHA}$ field mentioned above indicates if the star is active or not ($1=$ active and $0=$ inactive; described below). 
Both \cite{West2011} and \cite{Schmidt2015} classified stars as magnetically active or inactive according to the equivalent width of the $\halpha$ emission line. 
Stars were considered \textit{active} if they meet these four criteria: (1)~the signal-to-noise per pixel in the region near $\halpha$ is greater than 3, (2)~$\halpha$~EW~$> 0.75$~\r{A}, the detection threshold for SDSS spectra, (3)~$\halpha$~EW is larger than its uncertainty, and (4)~the peak height of $\halpha$ is greater than three times the noise level of the continuum region (measured by the standard deviation of the flux values). Stars that pass criterion (1) for signal-to-noise but do not pass the detection threshold are categorized as \textit{inactive}. Stars that do not pass the first criterion were classified as neither active nor inactive, and stars which pass all criteria but the second were classified as weakly active and not included in this work. 

Both the SDSS DR7 M dwarf spectroscopic catalog and the BUD catalog also include proper motions for the stars that were used as part of our cross-match procedure. The stars in the SDSS~DR7 M~dwarf catalog are all bright enough that proper motions were part of the \citet[USNO-B;][]{Munn2014} catalog, based on SDSS and USNO-B positions, and have mean uncertainties of only $5$~mas/yr.
The stars in the BUD sample, on the other hand, are too faint to be in the USNO-B, and their proper motions were calculated using positions from SDSS, 2MASS, and the Wide-Field Infrared Sky Explorer \citep[WISE;][]{Wright2010} by \citet[in prep]{schmidt2019} and have mean uncertainties of $20$~mas/yr. 

Lastly, both catalogs include radial velocities estimates with typical uncertainties of $\sim$~$7$~km/s. These were measured via cross-correlation of the ($R 1800-2200$) SDSS spectra to templates from \cite{Schmidt2014b} and \cite{Bochanski2007}. 

When compiling the MLSDSS sample, we modified several fields from the two input catalogs to make them as consistent as possible with each other: the photometry quality flag ($\texttt{GOODPHOT\_SDSS}$); the white dwarf-M Dwarf binary flag ($\texttt{WDM}$); and the photometry impacted by extinction.

The two input catalogs indicate good quality photometry using different methods: the SDSS DR7 M dwarfs spectroscopic catalog assigned a single quality flag $\texttt{GOODPHOT\_SDSS}=1$ or $0$ that depends on the quality of $r$, $i$ and $z$-band photometry ($r$-band extinction $<0.05$ magnitudes, and uncertainties $<0.05$ magnitudes) while the BUD catalog has a flag for each band (using a combination of SDSS flags and uncertainty cuts to select good photometry, see Section 3.1 in \citet{Schmidt2015} for more details). 
We applied the first convention to the BUD stars, assigning them a $\texttt{GOODPHOT\_SDSS}=1$ value if the $r$, $i$ and $z$-band were all good. 

Another difference between the two input catalogs is the $\texttt{WDM}$ flag, which indicates if the star is a white dwarf-M Dwarf binary ($\texttt{WDM}=1$ is a binary and $\texttt{WDM}=0$ is not). \cite{West2011} selected these pairs with the color cuts from \citet{Smolcic2004}: $u - g < 2$, $g - r > 0.3$, $r - i > 0.7$, $\sigma _{u,g,r,i} < 0.1$. The BUD catalog does not contain white dwarf-M Dwarf binaries \citep{Schmidt2015} so we added a $0$ in the $\texttt{WDM}$ column for all of these stars. 

Lastly, \cite{West2011} corrected all five SDSS magnitudes for dust extinction using the \cite{Schlegel1998} maps. \cite{Schmidt2015} did not apply the correction to the magnitudes and instead included the extinction correction as a field in the BUD catalog. We applied the extinction correction for the stars in the BUD catalog so that all of the included SDSS photometry in the MLSDSS sample is corrected for extinction.

\begin{figure}[t!]
\centering
\includegraphics[width=\linewidth]{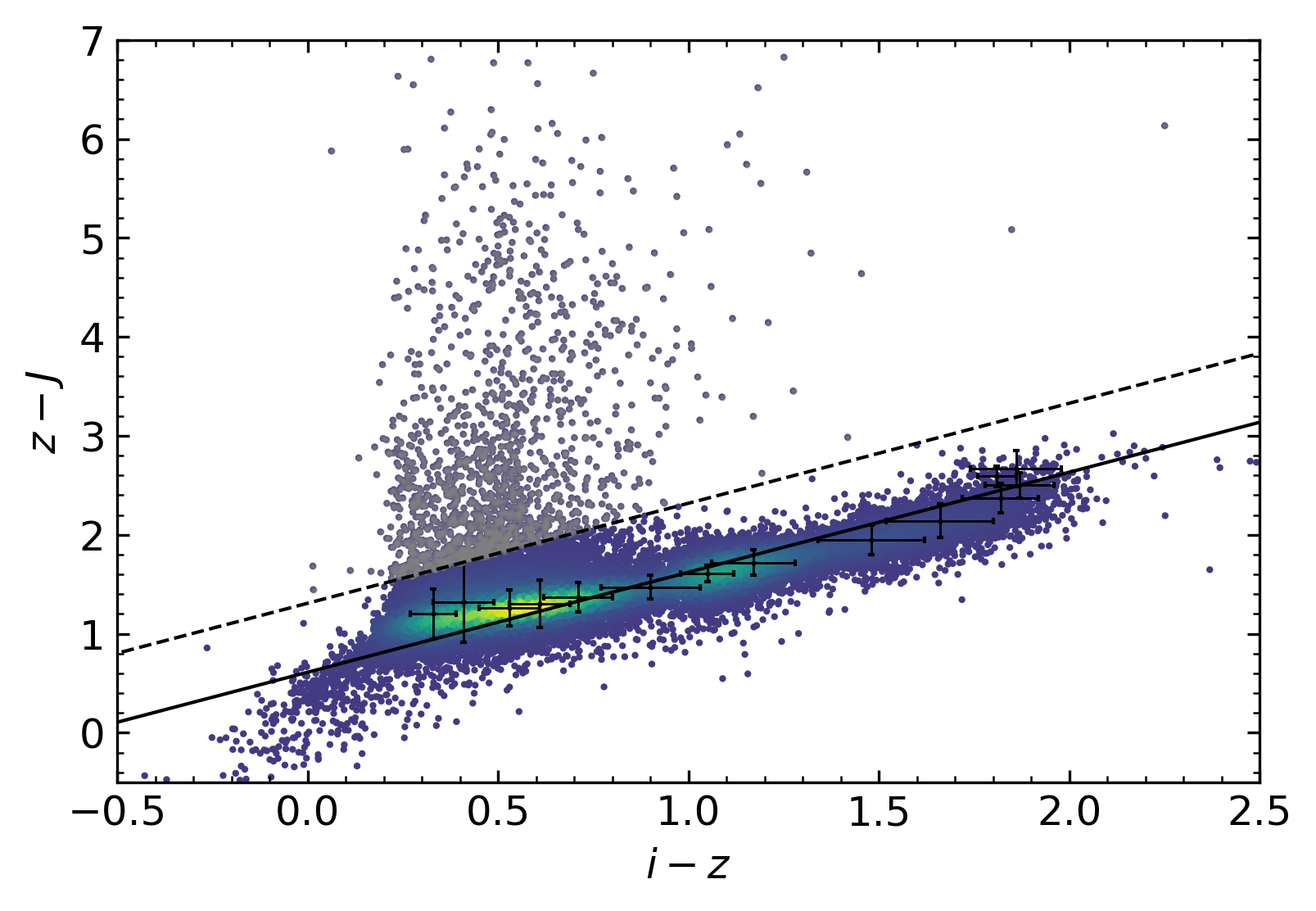}
\caption{Color-color plot, $(z-J)$ versus $(i-z)$, for the MLSDSS sample color-coded with the density map. Black crosses indicate the median values and standard deviation for confirmed M and L~dwarfs from \citet{West2011,Schmidt2015}. The linear fit to the median values is indicated with a solid black line. To perform the fit we used the errors in $(z-J)$ as weights. The very red $(z-J)$ colors for stars with bluer $(i-z)$ colors are likely to be due to mismatches between the MLSDSS sample and 2MASS. 
We removed the 2MASS information for the $1,494$ stars (gray points) with $(z-J)$ colors more than $3\sigma$ above the fit (dashed line).}
\label{fig:r_zvsr_k}
\end{figure} 

We found that some of the 2MASS photometry included in the DR7 M~dwarf catalog was incorrect due to mismatches.
This is demonstrated in Figure~\ref{fig:r_zvsr_k} where the $(z-J)$ versus $(i-z)$ color-color plot shows a significant scatter towards redder $(z-J)$ color, inconsistent with the $(i-z)$ colors and spectral types of the sample when compared to the median values and standard deviation for $(z-J)$ and $(i-z)$ colors of confirmed M and L~dwarfs \citep{West2011,Schmidt2015}.
The outliers do not have a low signal-to-noise ratio $J$-band photometry and have good SDSS photometry ($\texttt{GOODPHOT\_SDSS}=1$). They are also more common among fainter, bluer stars that are unlikely to be bright enough in 2MASS bands to have detections. It is therefore likely that the spurious colors are due to 2MASS mismatches with the SDSS source. 
We fit a line to the median values using the errors in $(z-J)$ as weights and removed the 2MASS information for the $1,494$ stars with $(z-J)$ colors more than $3\sigma$ above the fit, where $\sigma$ is the mean propagated error on the $(z-J)$ color. These $1,494$ stars remain in the MLSDSS sample, just without 2MASS photometry.

The final MLSDSS sample includes $74,216$ M and L~dwarfs with spectral types, SDSS $ugriz$ photometry, 2MASS $JHK_s$ photometry, $\halpha$ equivalent width and fractional luminosity ($L_{\rm H\alpha}/L_{\rm bol}$), an activity classification, proper motions, and radial velocities.

\subsection{Cross-match with \textit{Gaia} DR2}
\label{subsec:crossmatch}

We cross-matched the MLSDSS sample with \textit{Gaia} DR2 to obtain precise proper motions and parallaxes. First, we propagated the positions from the SDSS epoch (ranging from $1999$ to $2007$) to the \textit{Gaia} DR2 epoch ($2015.5$) using the proper motions in MLSDSS. 
Second, we queried the \textit{Gaia} Archive\footnote{http://gea.esac.esa.int/archive/} and selected all the objects within a radius of $5\arcsec$ of the 2015.5 position. We found that $98\%$ ($73,003$ stars) of MLSDSS stars have at least one match in \textit{Gaia} DR2. Of these, $8,269$ have between two and five matches within a $5\arcsec$ radius. To find a single best match, we propagated the position of each match back to the SDSS epoch using the \textit{Gaia} DR2 proper motion and kept only the closest match between the \textit{Gaia} position at the SDSS epoch and the SDSS position. We include a FITS table that contains the $73,003$~matches in our sample as a supplementary file. In Table~\ref{table:sampleMLSDSSGaiaDR2} we list the parameters in our sample and in the FITS table.

For this paper, we want a high fidelity sample with a minimum of mismatches. We found that a $1\arcsec$ separation between SDSS (RA, DEC), and Gaia (RA, DEC) propagated backwards to the SDSS epoch using the proper motions from \textit{Gaia}, provides a reasonable balance between sample size and crossmatch reliability. A total of $67,573$ stars ($91\% $ of MLSDSS) have \textit{Gaia} DR2 matches and a separation less than or equal to~$1\arcsec$. These are indicated in Table~\ref{table:sampleMLSDSSGaiaDR2} with the $\texttt{GOODMATCH}$ flag ($\texttt{GOODMATCH} = 1$ or $0$). The analysis in this paper is based on these $67,573$ objects with matches ($\texttt{GOODMATCH} = 1$) and we call this the ``MLSDSS-GaiaDR2 sample''. 

\begin{figure}[ht!]
\begin{center}
\includegraphics[width=\linewidth]{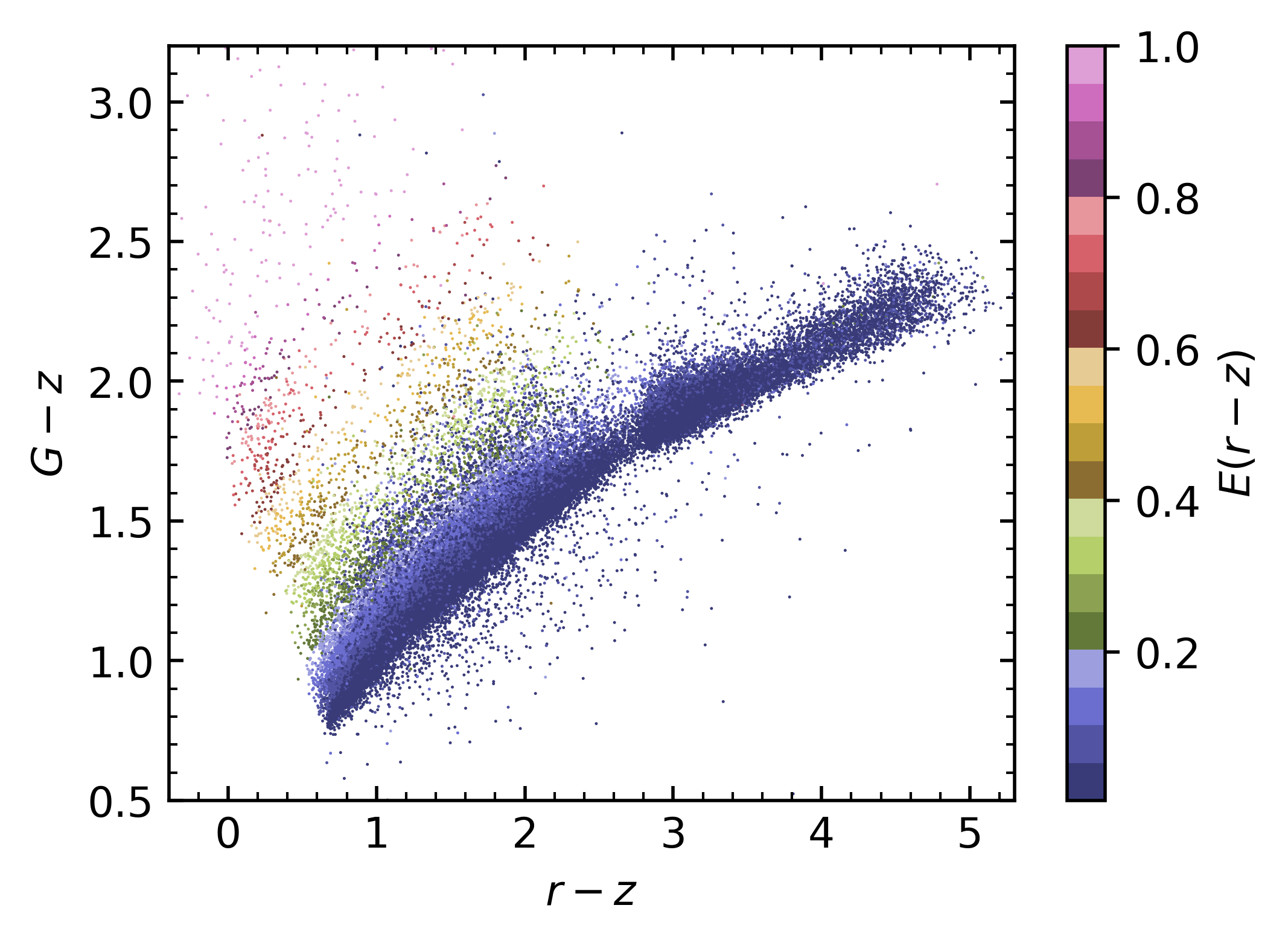}
\caption{$(G-z)$ as a function of $(r-z)$, with color-coding indicating the SDSS extinction value $E(r-z)$ using the \citet{Schlegel1998} dust map. Each star shown is classified both as a good match between the MLSDSS sample and \textit{Gaia} DR2 ($\texttt{GOODMATCH} = 1$) and has good SDSS photometry ($\texttt{GOODPHOT\_SDSS} = 1$), as discussed in Section~\ref{sec:MLSDSS}. The stars that fall along the color-color locus are easily verified as good matches. The stars that scatter towards redder $(G-z)$ color are those with high extinction, indicating that their position off the color-color locus is likely due to the lack of extinction correction to the $G$ band rather than mismatches in the catalog. 
}
\label{fig:G_zvsr_z}
\end{center}
\end{figure}

To check the goodness of the cross-matching, we examine a color-color plot of the MLDSS-GaiaDR2 sample in Figure \ref{fig:G_zvsr_z}. Nearly all of the stars fall along the expected locus, and the $\sim 800$ that fall off the locus are those with a high extinction correction to the SDSS magnitudes. The scatter in the color-color space is due to the lack of extinction correction applied to the $G$~magnitude, and does not indicate mismatches. Extinction corrections were not applied to the \textit{Gaia} photometry in our sample because the extinction coefficients provided by the collaboration were calibrated for ${\rm T_{eff}} > 3500$~K and are not valid for low mass stars \citep{Collaboration2018a}.

\clearpage

\startlongtable
\begin{deluxetable*}{ccl}
\tabletypesize{\scriptsize}
\tablecaption{Columns in the MLSDSS-GaiaDR2 sample. \label{table:sampleMLSDSSGaiaDR2}}
\tablehead{\colhead{name} & \colhead{units} & \colhead{description}}
\startdata
$\texttt{MJD}$ & d & Modified julian date from SDSS \\
$\texttt{PLATE}$ &  & Plate number from SDSS \\
$\texttt{FIBER}$ &  & Fiber number from SDSS \\
$\texttt{solution\_id}$ &  & Gaia DR2 Solution Identifier \\
$\texttt{designation}$ &  & Unique Gaia source designation (unique across all Data Releases) \\
$\texttt{source\_id}$ &  & Unique Gaia source identifier (unique within DR2) \\
$\texttt{ref\_epoch\_gaia}$ & yr & Reference epoch from Gaia DR2 \\
$\texttt{SPT}$ &  & Spectral Type \\
$\texttt{RA}$ & deg & Right ascension in Gaia DR2 epoch \\
$\texttt{RA\_ERR}$ & mas & Standard error of right ascension in Gaia DR2 \\
$\texttt{DEC}$ & deg & Declination in Gaia DR2 epoch \\
$\texttt{DEC\_ERR}$ & mas & Standard error of DEC in Gaia DR2 \\
$\texttt{PMRA}$ & mas/yr & Proper motion in RA direction in Gaia DR2 \\
$\texttt{PMRA\_ERR}$ & mas/yr & Standard error of proper motion in RA direction in Gaia DR2 \\
$\texttt{PMDEC}$ & mas/yr & Proper motion in DEC direction in Gaia DR2 \\
$\texttt{PMDEC\_ERR}$ & mas/yr & Standard error of proper motion in DEC direction in Gaia DR2 \\
$\texttt{RV}$ & km/s & Radial velocity from MLSDSS \\
$\texttt{RV\_ERR}$ & km/s & Radial velocity error from MLSDSS \\
$\texttt{RA\_SDSS}$ & deg & Right ascension in SDSS photometric object \\
$\texttt{DEC\_SDSS}$ & deg & Declination in SDSS photometric object \\
$\texttt{PSFMAG}$ & mag & SDSS photometry $ugriz$-bands \\
$\texttt{PSFMAG\_ERR}$ & mag & SDSS photometry $ugriz$-bands errors \\
$\texttt{GOODPHOT\_SDSS}$ &  & Good photometry flag for SDSS riz-bands \\
$\texttt{EXTINCTION}$ & mag & Extinction coorection for $ugriz$-bands ($A_{\rm u}$, $A_{\rm g}$, $A_{\rm r}$, $A_{\rm i}$, $A_{\rm z}$) \\
$\texttt{PMRA\_SDSS}$ & mas/yr & Proper motion in RA direction in MLSDSS \\
$\texttt{PMRA\_ERR\_SDSS}$ & mas/yr & Proper motion error in RA direction in MLSDSS \\
$\texttt{PMDEC\_SDSS}$ & mas/yr & Proper motion in DEC direction in MLSDSS \\
$\texttt{PMDEC\_ERR\_SDSS}$ & mas/yr & Proper motion error in DEC direction in MLSDSS \\
$\texttt{GOODPM\_SDSS}$ &  & Good proper motion flag for MLSDSS (1=good proper motion) \\
$\texttt{MJD\_2MASS}$ & d & Modified julian date from 2MASS \\
$\texttt{RA\_2MASS}$ & deg & Right ascension in 2MASS \\
$\texttt{DEC\_2MASS}$ & deg & Declination in 2MASS \\
$\texttt{MAG\_2MASS}$ & mag & 2MASS photometry $JHK$-bands \\
$\texttt{MAG\_ERR\_2MASS}$ & mag & 2MASS photometry $JHK$-bands error \\
$\texttt{ACTHA}$ &  & Active flag (1=active, 0=inactive) \\
$\texttt{EWHA}$ & Angstrom & Equivalent width $\halpha$ \\
$\texttt{EWHA\_ERR}$ & Angstrom & Equivalent width $\halpha$ error \\
$\texttt{LHALBOL}$ &  & Fractional $\halpha$ luminosity \\
$\texttt{LHALBOL\_ERR}$ &  & Fractional $\halpha$ luminosity error \\
$\texttt{GOODMATCH}$ &  & Good matches with Gaia DR2 (1=good, 0=probable mismatch) \\
$\texttt{parallax}$ & mas & Parallax in Gaia DR2 \\
$\texttt{parallax\_error}$ & mas & Standard error of parallax in Gaia DR2 \\
$\texttt{astrometric\_n\_good\_obs\_al}$ &  & Number of good observations AL \\
$\texttt{astrometric\_chi2\_al}$ &  & AL chi-square value \\
$\texttt{visibility\_periods\_used}$ &  & Number of visibility periods used in Astrometric solution \\
$\texttt{phot\_g\_mean\_flux}$ & electron/s & $G$-band mean flux \\
$\texttt{phot\_g\_mean\_flux\_error}$ & electron/s & Error on $G$-band mean flux \\
$\texttt{phot\_g\_mean\_flux\_over\_error}$ & electron/s & $G$-band mean flux divided by its error \\
$\texttt{phot\_g\_mean\_mag}$ & mag & $G$-band mean magnitude \\
$\texttt{phot\_bp\_mean\_flux}$ & electron/s & Integrated $\gbp$ mean flux \\
$\texttt{phot\_bp\_mean\_flux\_error}$ & electron/s & Error on the integrated $\gbp$ mean flux \\
$\texttt{phot\_bp\_mean\_flux\_over\_error}$ & electron/s & Integrated $\gbp$ mean flux divided by its error \\
$\texttt{phot\_bp\_mean\_mag}$ & mag & Integrated $\gbp$ mean magnitude \\
$\texttt{phot\_rp\_mean\_flux}$ & electron/s & Integrated $\grp$ mean flux \\
$\texttt{phot\_rp\_mean\_flux\_error}$ & electron/s & Error on the integrated $\grp$ mean flux \\
$\texttt{phot\_rp\_mean\_flux\_over\_error}$ & electron/s & Integrated $\grp$ mean flux divided by its error \\
$\texttt{phot\_rp\_mean\_mag}$ & mag & Integrated $\grp$ mean magnitude \\
$\texttt{phot\_bp\_rp\_excess\_factor}$ &  & BP/RP excess factor \\
$\texttt{r\_est}$ & pc & B-J estimated distance \\
$\texttt{r\_lo}$ & pc & B-J lower bound on the confidence interval of the estimated distance \\
$\texttt{r\_hi}$ & pc & B-J upper bound on the confidence interval of the estimated distance \\
$\texttt{r\_len}$ & pc & B-J length scale used in the prior for the distance estimation \\
$\texttt{V\_R}$ & km/s & Mean radial component of the velocity \\
$\texttt{V\_R\_ERR}$ & km/s & Standard deviation of radial component of the velocity \\
$\texttt{V\_T}$ & km/s & Mean tangential component of the velocity \\
$\texttt{V\_T\_ERR}$ & km/s & Standard deviation of tangential component of the velocity \\
$\texttt{V\_Z}$ & km/s & Mean vertical component of the velocity \\
$\texttt{V\_Z\_ERR}$ & km/s & Standard deviation of vertical component of the velocity \\
$\texttt{J\_Z}$ & kpc km/s & Median vertical action \\
$\texttt{J\_Z\_16per}$ & kpc km/s & 16th percentile vertical action \\
$\texttt{J\_Z\_64per}$ & kpc km/s & 64th percentile vertical action \\
$\texttt{WDM}$ &  & White dwarf-M dwarf binary flag (1=binary, 0=not binary) \\
$\texttt{astrometric\_sample}$ &  & Astrometric subsample (1=good astrometry, 0=bad astrometry) \\
$\texttt{photometric\_sample\_subg}$ &  & Sub G subsample (1=goodphot, 2=outlier, 0=badphot) \\
$\texttt{photometric\_sample\_subred}$ &  & Sub Red subsample (1=goodphot, 2=outlier, 0=badphot) \\
$\texttt{photometric\_sample\_submix}$ &  & Sub Mix subsample (1=goodphot, 2=outlier, 0=badphot)
\enddata
\tablecomments{Columns of the MLSDSS-GaiaDR2 sample of $74,216$ M and L dwarfs, including the $73,003$ matches with Gaia DR2. The table is available as a FITS table as a supplementary file. We include name of the columns, units and a brief description.}
\end{deluxetable*}


\clearpage

\subsection{Quality Cuts}
\label{subsec:qualitycuts}

To ensure the cleanest possible sample of \textit{Gaia} DR2 M and L dwarfs, we investigated optimal quality cuts for the photometric and astrometric data. The quality cuts in the \textit{Gaia} Papers \citep[e.g.][]{Lindegren2018, Evans2018,Arenou2018,Collaboration2018a} are so conservative that they remove good quality data for faint, red stars at the end of the main sequence. As as result, we modified the suggested cuts to adapt them for M and L dwarfs, the faintest stars in \textit{Gaia} DR2. We describe these cuts in the following subsections. 

\subsubsection{Astrometric Quality Cuts}
\label{subsubsec:astrometriccuts}

The quality of the five-parameter solution (\texttt{ra}, \texttt{dec}, \texttt{pmra}, \texttt{pmdec} and \texttt{parallax}) given by \textit{Gaia} DR2 depends on factors such as the magnitude of the source, the number of observations per source, neighboring sources, and the type of source \citep{LL:LL-124}. We describe below how we defined astrometric cuts for the MLSDSS-GaiaDR2 sample to obtain the best quality five-parameter solution. The astrometric cuts we used to clean the MLSDSS-GaiaDR2 sample are summarized in Table~\ref{table:flags} and described below.

\begin{deluxetable*}{cccc}[ht!]
\tablecaption{Summary of Astrometric Quality Cuts. \label{table:flags}}
\tablehead{\colhead{Flag} & \colhead{Cut} & \colhead{$N$ removed by} &  \colhead{$N$ after}\\
\colhead{} & \colhead{} & \colhead{single cut} & \colhead{cumulative cuts}}
\startdata
& & & start=67,573 \\
\texttt{PE} &  $\texttt{parallax\_over\_error} > 10$   & 40,801 & $26,772$ \\ 
\texttt{VP} & $\texttt{visibility\_periods\_used} > 8$ & 8,166 & $24,589$ \\ 
UWE & UWE$< 1.2 \times \max(1.4, \exp(-0.2(G - 19.5)))$ & 2,582 & $23,842$\\ 
\enddata
\tablecomments{Astrometric quality cuts applied to the MLSDSS-GaiaDR2 sample. The \textit{Flag} column contains the name of the cut we use in this paper; \textit{Cut} indicates the name of the column in the catalog and the criterion applied; \textit{N removed by Single Cut} shows the number of stars removed by only that cut; and \textit{N after Cumulative cuts} shows the number of stars left after applying that cut and the ones listed above it. Objects included in the astrometric sample are indicated with the flag $\texttt{astrometric\_sample}=1$.}
\end{deluxetable*}

To ensure accurate parallaxes (mean uncertainty $\sim$~$0.2$~mas) we applied the quality cut suggested by \cite{Lindegren2018}: $\texttt{parallax\_over\_error} > 10$ (abbreviated as \texttt{PE} from here on). This cut conservatively removes poor astrometric solutions and reduces our sample by $60\%$, removing $40,801$ stars.

The number of \textit{Gaia} observations included in each astrometric solution is an indicator of reliable astrometric data and is indicated in the $\texttt{visibility\_periods\_used}$ field, abbreviated as \texttt{VP} from here on. 
As suggested in \cite{Collaboration2018a}, we selected stars with $\texttt{VP} > 8$ to restrict our sample to stars with enough observations to produce reliable astrometric solutions. This removes $8,166$ stars from the original MLSDSS-GaiaDR2 sample, leaving $24,589$ stars when applied after the \texttt{PE} cut (see Table~\ref{table:flags}).

To remove poor astrometric solutions generated by binary stars and double stars, we also applied an astrometric cut based on the residual of the fit of the single star astrometric solution. 
The ``unit weight error'' (UWE) is a reduced $\chi^2$ statistic and reflects the goodness of fit \citep{Arenou2018,Lindegren2018}. 
The square of the UWE is calculated as 

\begin{equation}\label{eq:ucut}
\textrm{UW}\textrm{E}^2 = \chi^2/\nu=\frac{\texttt{astrometric\_chi2\_al}}{\texttt{astrometric\_n\_good\_obs\_al}-5}    
\end{equation}

\noindent where $\chi ^2=\texttt{astrometric\_chi2\_al}$ and $\nu =N-5$ is the degree of freedom where $N=\texttt{astrometric\_n\_good\_obs\_al}$ is the total number of good observations of the source.
\cite{Lindegren2018} found that a good astrometric solution corresponds to UWE $\sim 1$ and suggest a cut: 

\begin{equation}
\textrm{UWE} < 1.2 \times \max(1, \exp(-0.2(G - 19.5))).  \label{eq:oldAF}
\end{equation}

\begin{figure}[ht!]
\begin{center}
\includegraphics[width=\linewidth]{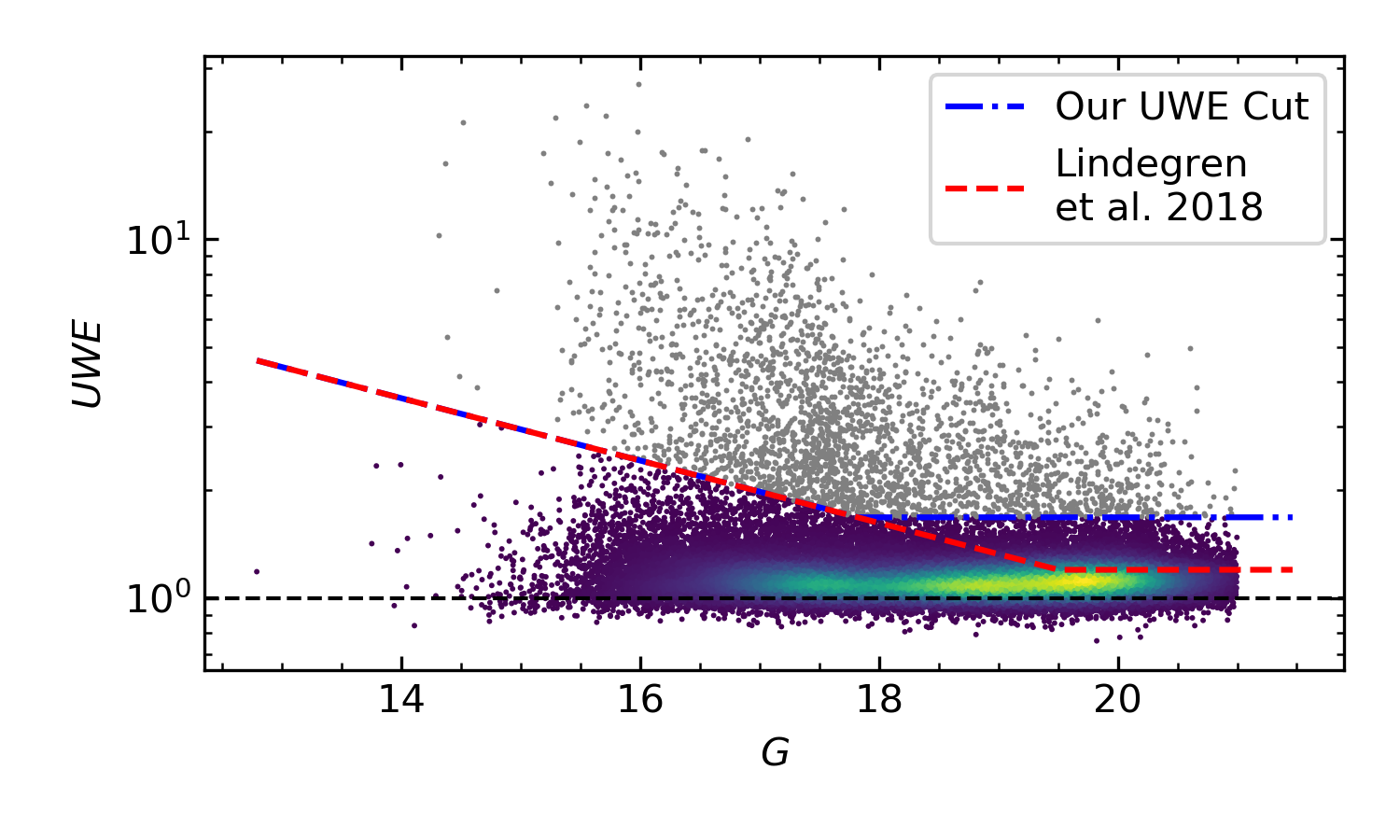}
\caption{Unit weight error (UWE, defined in Equation \ref{eq:ucut}) as a function of $G$ magnitude for the MLSDSS-GaiaDR2 sample color-coded with the density map. The \cite{Lindegren2018} cut (red dashed line) defined in Equation \ref{eq:oldAF} removes most of the faintest stars (later Ms and Ls) with $G\ge18$ although they have UWE$\sim 1$. We defined a new cut (blue dashed-dotted line) in Equation \ref{eq:newAF}, that both removes stars with a bad astrometric solution and keeps the fainter stars. Stars retained after this cut are bellow the blue dashed-dotted line.} 
\label{fig:astrometric_chi}.
\end{center}
\end{figure}

We show UWE as a function of $G$ magnitude in Figure~\ref{fig:astrometric_chi}. The cut suggested by \cite{Lindegren2018}, shown as a red dashed line, removes a high number of faint stars ($G>18$) even though they have a good astrometric fit (UWE$\sim 1$). We wanted to retain faint stars for our sample of M and L dwarfs and future analysis, so we defined a new cut and increase the maximum UWE tolerance for faint stars from 1.2 to 1.68: 

\begin{equation}
\textrm{UWE} < 1.2 \times \max(1.4, \exp(-0.2(G - 19.5)))   \label{eq:newAF}
\end{equation}

\noindent represented in a blue dashed-dotted line in Figure \ref{fig:astrometric_chi}. This new cut matches the \cite{Lindegren2018} through $G=18$ and includes an extra $7,132$ stars with $G>18$ also having a good astrometric solution.
Applying this cut removes $2,582$ stars from the original MLSDSS-GaiaDR2 sample, leaving $23,842$ stars when applied after the \texttt{PE} and \texttt{VP} cuts (see Table~\ref{table:flags}).

An alternative to this cut is described in the Technical Note by \citet{LL:LL-124}, where they define a new quantity called the re-normalized UWE, or RUWE. Because the UWE is necessarily dependent on the color and magnitude of each star, the RUWE is designed to make a quality cut in the data that is relatively complete in color and magnitude. This is calculated by dividing UWE by a different normalization factor for each color and magnitude bin, which accounts for the fraction of good and bad data in each bin. We did not intend our sample to be complete in color and/or magnitude, and applying a cut on the RUWE removes $\sim 1,000$ more stars than Equation \ref{eq:newAF}, so we did not use RUWE in our quality cuts.

\begin{figure}[ht!]
\begin{center}
\includegraphics[width=\linewidth]{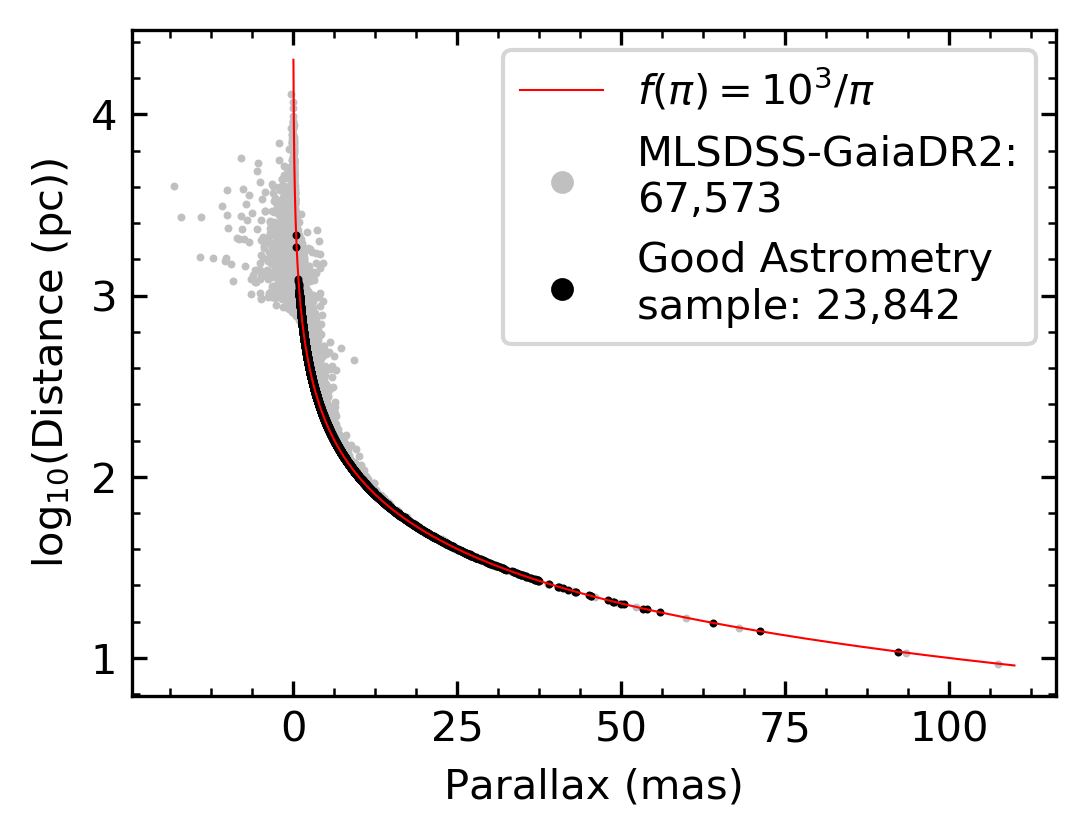}
\caption{Log of distance calculated by \cite{Bailer-Jones2018} as a function of the parallaxes measured in \textit{Gaia} DR2. The MLSDSS-GaiaDR2 sample before our astrometric cuts (gray dots) is compared to the astrometric sample (black dots). The formula relating distance to parallax (solid red line) shows excellent agreement with the astrometric sample, validating the astrometric cuts applied to the MLSDSS-GaiaDR2 sample.} 
\label{fig:astrometric_cuts}.
\end{center}
\end{figure}

Once the three astrometric cuts summarized in Table \ref{table:flags} are applied, the MLSDSS-GaiaDR2 sample contains $23,842$ stars with good astrometry. As a way of verifying our astrometric cuts, we cross-matched MLSDSS-GaiaDR2 with the \cite{Bailer-Jones2018} catalog that uses an inference procedure to account for the non linearity of $1/\pi$ for computing distances. If our astrometric cuts are valid, the distances calculated as $1/\pi$ with MLSDSS-GaiaDR2 should be the same as in the \cite{Bailer-Jones2018} catalog. We plot parallaxes from our sample against \cite{Bailer-Jones2018} distances in Figure \ref{fig:astrometric_cuts} and confirm that the $23,842$ parallaxes selected by our cuts follow the formula relating parallax and distance, $d=10^3/\pi$, where $\pi$ is the parallax in mas. This indicates that the quality of the astrometry in our final sample is excellent. Objects included in the astrometric sample are indicated with the flag $\texttt{astrometric\_sample}=1$ in the FITS file.

\subsubsection{Photometric Quality Cuts for Gaia bands}
\label{subsubsec:photcuts}

Given that the MLSDSS-GaiaDR2 sample contains predominantly faint and red stars, we also implemented several photometric cuts described below to ensure a sample without contamination and suitable for detailed color analysis. The cuts and the resulting subsamples are summarized in Table~\ref{table:flagsphot} and described below.

\begin{deluxetable*}{ccc}[ht!]
\tablecaption{Summary of Photometric Quality Cuts. \label{table:flagsphot}}
\tablehead{\colhead{Subsample} & \colhead{Cut expression} & \colhead{$N$ Stars} }
\startdata
& & start = 23,842 \\
(1) Sub G & astrometric cuts   &  $22,706$ \\ 
 & + $\texttt{phot\_bp\_rp\_excess\_factor} < 1.3+0.06 \times( \gbp - \grp)^2$ & \\
 &  only when $\texttt{phot\_bp\_mean\_flux\_over\_error} > 10$ & \\ 
(2) Sub Red & (1) + $\texttt{phot\_rp\_mean\_flux\_over\_error}>10$ & $22,373$\\
(3) Sub Mix & (2) + $\texttt{phot\_bp\_mean\_flux\_over\_error}>10$ & $16,527$ \\
\enddata
\tablecomments{Photometric quality cuts applied to the good astrometry sample (Section \ref{subsubsec:astrometriccuts}). The column \textit{Subsample} indicates the name of the subsample used in this paper; the \textit{Cut expression} indicates the name of the column in the catalog and the cuts that were made over that column; and \textit{N Stars} indicates the number of stars in each subsample. Objects in the subsamples are indicated with the flags, $\texttt{photometric\_sample\_subg} = 1$, $\texttt{photometric\_sample\_subred} = 1$ and $\texttt{photometric\_sample\_submix} = 1$, respectively.}
\end{deluxetable*}

\begin{figure*}[ht!]
\begin{center}
\includegraphics[width=\linewidth]{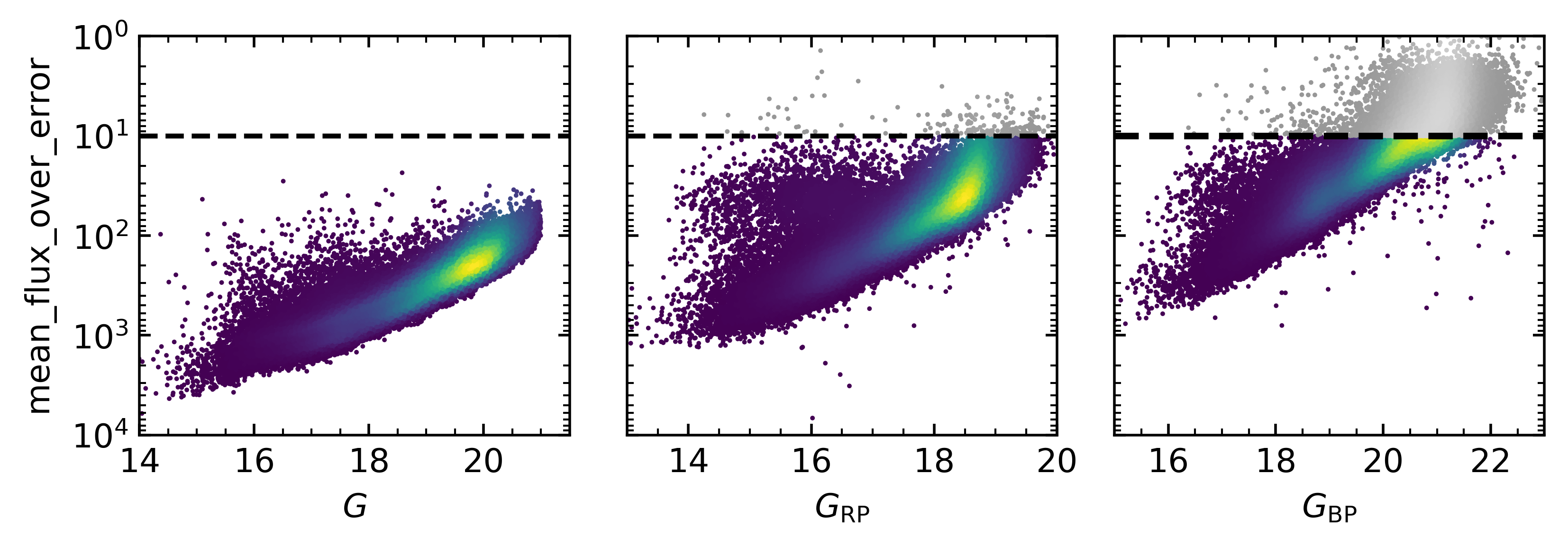}
\caption{Flux SNR ($\texttt{mean\_flux\_over\_error}$) for each Gaia band as a function of magnitude color-coded with the density map. The black dashed line is the limit of $10\%$ SNR suggested for the red, $\grp$ and blue, $\gbp$ bands in \cite{Lindegren2018}. The SNR limit applied to the $G$ band does not remove stars, while the $\gbp$ band removes the most stars, in particular faint ones. Note that not every star that has bad blue, $\gbp$ photometry has bad red, $\grp$ photometry or vice versa.} 
\label{fig:flux_over_error}.
\end{center}
\end{figure*}

To ensure accurate \textit{Gaia} photometry (signal-to-noise ratio $> 10$) we applied cuts based on the signal-to-noise ratio (SNR) for the flux in the three \textit{Gaia} bands, $G$ ($[330, 1050]$ nm), $\gbp$ ($[330, 680]$ nm), and $\grp$ ($[630, 1050]$ nm). We show the mean flux over error for these three bands in Figure~\ref{fig:flux_over_error}. \textit{Gaia} DR2 contains a column with the SNR value for each band. \cite{Lindegren2018}, in their Appendix~C, suggest combining the cuts over SNR for the blue and red band ($\gbp$ and $\grp$ respectively): $\texttt{phot\_bp\_mean\_flux\_over\_error} > 10$ and $\texttt{phot\_rp\_mean\_flux\_over\_error} > 10$. The suggested cut for the blue \textit{Gaia} band removes a significant number of stars from the MLSDSS-GaiaDR2 sample, while the same cut in the red band $\grp$ only removes a handful of stars. This is expected because M and L dwarfs emit most of their flux at red wavelengths, so they are faint in the blue band. If we follow the suggestion made by \cite{Lindegren2018} and combine the cuts for the red and blue bands, we would remove $5,846$ stars that have SNR$<$10 in $\gbp$, but SNR$\ge$10 in $\grp$. Furthermore, all the stars in the MLSDSS-GaiaDR2 sample have SNR$\ge$10 in the $G$ band as we show in the left panel in Figure \ref{fig:flux_over_error}. In particular, the $5,846$ stars that have low-quality blue photometry, $\gbp$ but high-quality red photometry, $\grp$, have also good $G$ photometry. To maximize the number of stars available for each band with high SNR photometry, we created three subsamples: in the first subsample (Sub G) we did not apply any SNR cuts, only the $G$ photometry is necessarily SNR$\ge$10; in the second subsample (Sub Red) we applied the SNR cut in the red band ($\grp$), resulting in good $G$ and $\grp$ photometry; and in the third subsample (Sub Mix) we applied the SNR cut to both the red and blue bands so, therefore, it contains SNR$\ge$10 photometry in $G$, $\grp$ and $\gbp$ bands. The summary of these subsamples is presented in Table \ref{table:flagsphot}. Objects in the subsamples are indicated with the flags, $\texttt{photometric\_sample\_subg} = 1$, $\texttt{photometric\_sample\_subred} = 1$ and $\texttt{photometric\_sample\_submix} = 1$, respectively.

\begin{figure*}[ht!]
\begin{center}
\includegraphics[width=\linewidth]{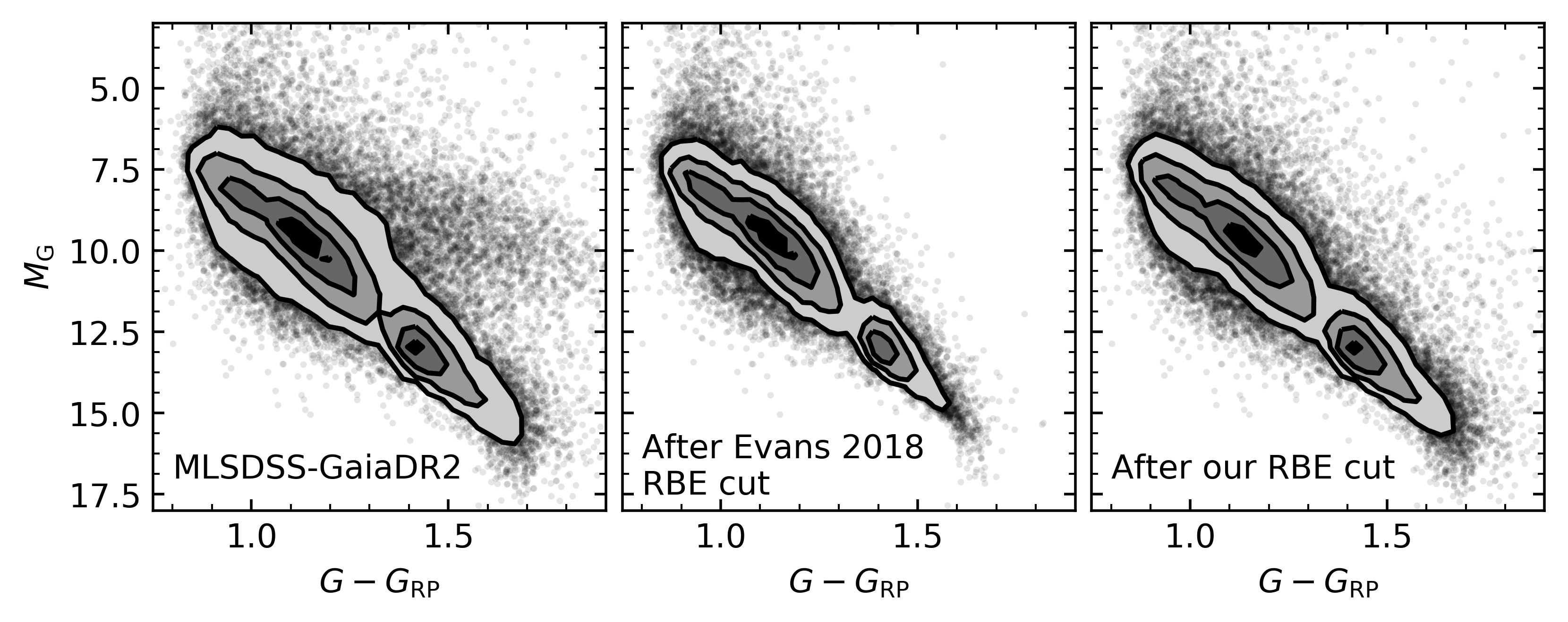}
\caption{Absolute magnitude $M_G$ versus $(G-\grp)$ color with the MLSDSS-GaiaDR2 sample in gray contours. On the left panel is the sample with no cuts. In the middle panel we applied the cut on BP/RP flux excess factor suggested in \cite{Evans2018} (Equation \ref{eq:oldrbe}) and on the right panel, the stars after applying the same cut with the condition $\texttt{phot\_bp\_rp\_excess\_factor} > 10$ (our cut). Our cut both removes contaminated photometry data and keeps fainter stars as can be seen by comparing the contours in the middle and right panels.} 
\label{fig:old_new_excessflag}.
\end{center}
\end{figure*}

The last source of photometric inaccuracy relevant to our sample is  contamination generated by neighbouring sources. As explained in \cite{Evans2018}, the wavelength ranges of the $\grp$ and $\gbp$ passbands overlap slightly. Therefore, the excess ratio defined as the flux ratio $C = ( I_{\rm RP}+I_{\rm BP})/I_{\rm G}$, where $I$ is the flux in the band indicated by the subindex, should be only slightly greater than~$1$. This quantity is indicated in \textit{Gaia} DR2 in the column $\texttt{phot\_bp\_rp\_excess\_factor}$. \cite{Evans2018} and \cite{Arenou2018} suggest the following criteria to select stars with uncontaminated photometry:
\begin{equation}\label{eq:oldrbe}
\texttt{phot\_bp\_rp\_}\texttt{excess\_factor} < 1.3+0.06 \times(\gbp - \grp)^2  
\end{equation}
The cut in Equation \ref{eq:oldrbe} selects the stars for which the excess factor $C$, is close to $1$. However, it depends on accurate $\gbp$ photometry, which is not available for our faint, red stars. Accordingly, the excess factor increases for fainter stars as a function of the three bands. We examine this cut in the color--magnitude diagram shown in Figure~\ref{fig:old_new_excessflag}. If we apply the cut suggested by \cite{Evans2018} to the MLSDSS-GaiaDR2 sample, it removes the spurious data shown in Figure \ref{fig:old_new_excessflag} (left panel has no cuts and the middle panel has these cuts applied). 
However, it also removes stars at the bottom of the main sequence that we are interested in keeping for future analysis because they have good $G$ and $\grp$ photometry. To reduce the number of high quality stars being eliminated for conservative $I_{\rm BP}$ values that generate a large excess factor, we applied the cut on the excess factor in Equation~\ref{eq:oldrbe} only when the blue photometry is good ($\texttt{phot\_bp\_mean\_flux\_over\_error} > 10$, abbreviated as \texttt{RBE} cut hereon). After adding this condition, the new cut to MLSDSS-GaiaDR2 removes significantly fewer main sequence stars (right panel of Figure \ref{fig:old_new_excessflag}). We applied this cut over the excess factor for the three subsamples as indicated in Table \ref{table:flagsphot}.

\begin{figure}[t]
\begin{center}
\includegraphics[width=\linewidth]{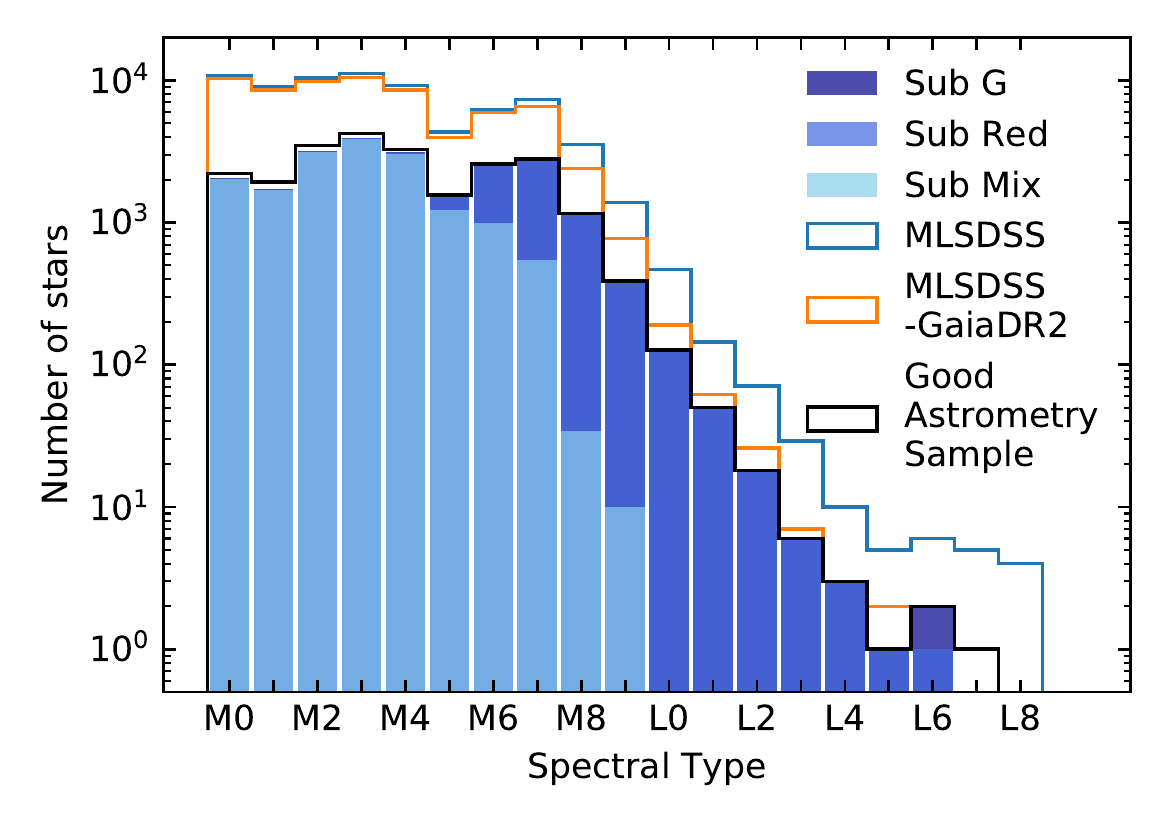}
\caption{Spectral type distribution for each of the three subsamples in Table \ref{table:flagsphot}. As described in Section~\ref{subsubsec:photcuts}, the Sub G photometric subsample has good $G$ photometry, the Sub Red sample has good $G$ and $\grp$ photometry and the Sub Mix sample has stars with good $G$, $\grp$ and $\gbp$. There is not a significant difference between the spectral type distribution of the Sub G and Sub Red sample. However, for the Sub Mix sample most of late M dwarfs and all L dwarfs were removed by the quality cut over the blue \textit{Gaia} band, $\gbp$. We added, for reference, the spectral type distribution of MLSDSS, MLSDSS-GaiaDR2 and the good astrometry sample. Later spectral types are not in the MLSDSS-GaiaDR2 sample, in comparison with MLSDSS, because they are too faint for \textit{Gaia} and we did not find a match.} 
\label{fig:cuts_hist}.
\end{center}
\end{figure}

The final spectral type distribution for the three photometric subsamples is shown in Figure \ref{fig:cuts_hist} compared to the entire MLSDSS, MLSDSS-GaiaDR2, and astrometric samples. Comparing the MLSDSS and the MLSDSS-GaiaDR2 samples, a significant difference can be observed in the number of late M and L dwarfs because \textit{Gaia} DR2 does not contain the faintest stars, so we could not find a match for all MLSDSS objects. The Sub G and Sub Red subsamples are similar to the astrometric sample because high quality G~photometry is necessary for the astrometric sample, and only a few stars have low SNR in $\grp$. The distribution changes significantly for the Sub Mix subsample because the SNR cut for $\gbp$ removed all the L dwarfs and many late-M dwarfs.

\begin{figure*}[t!]
\begin{center}
\includegraphics[width=\linewidth]{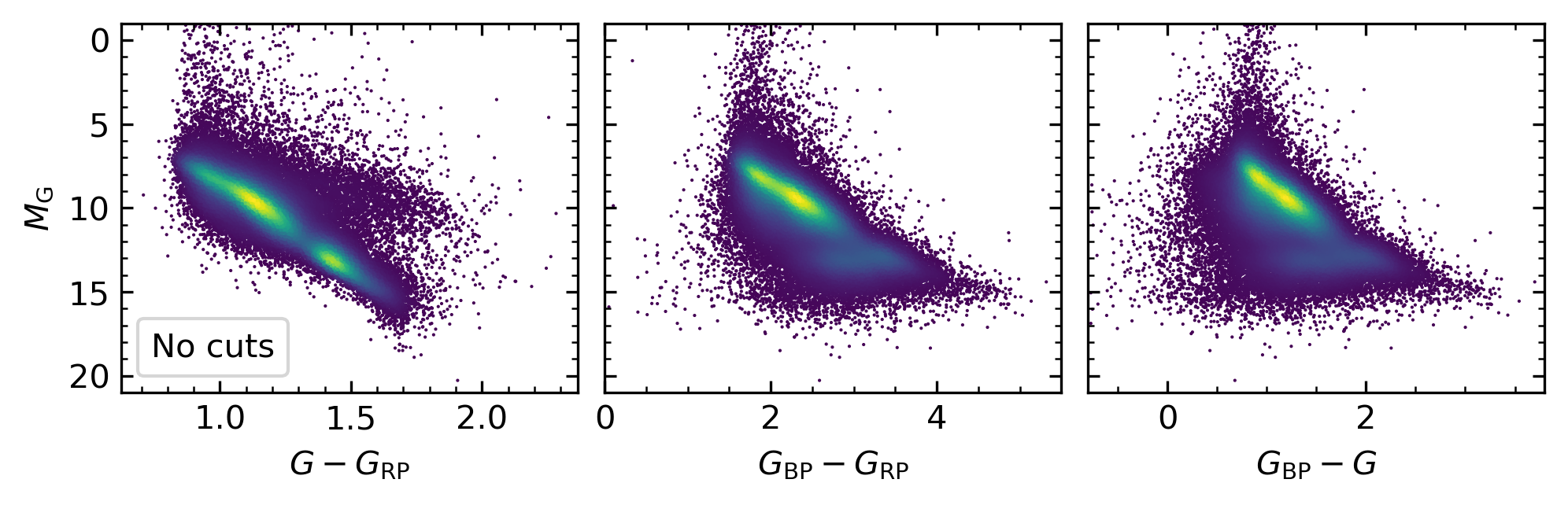}
\includegraphics[width=\linewidth]{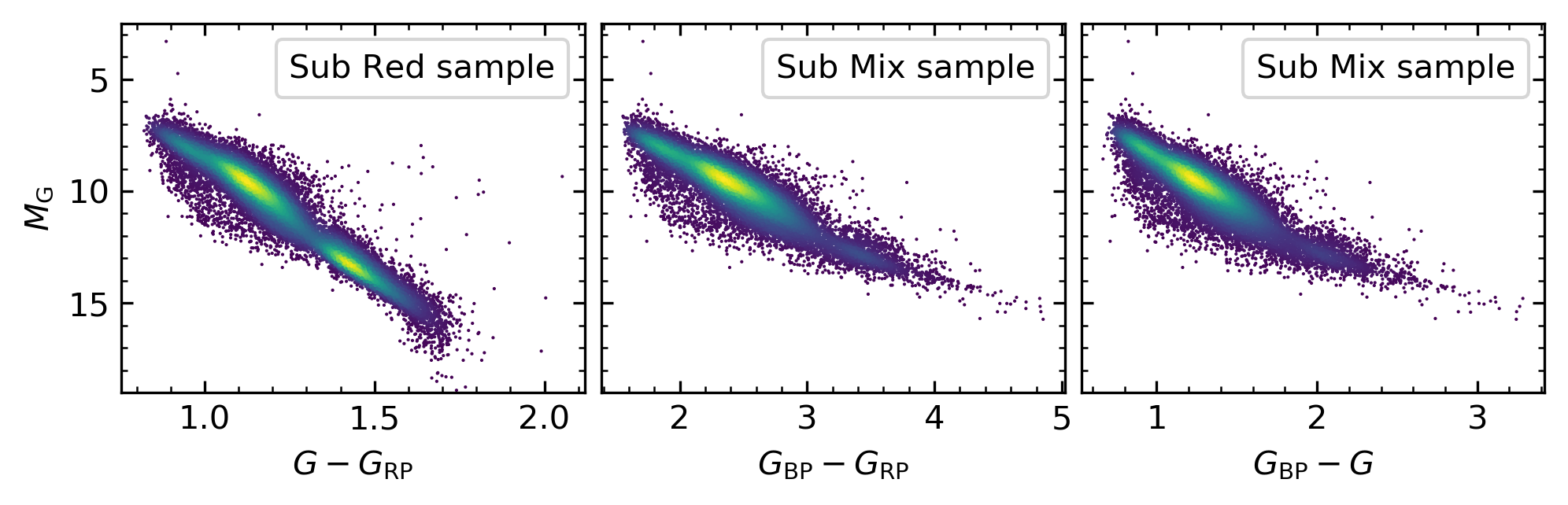}
\caption{Absolute magnitude $M_G$ versus \textit{Gaia} DR2 color ($G-\grp$, $\gbp-\grp$ and $\gbp-G$) for the original MLSDSS-GaiaDR2 sample (without cuts, top panels) and after the astrometric and photometric quality cuts (bottom panels). The color coding shows the density of sources, where yellow areas are more dense and purple ones, less dense. Two different photometric subsamples were used in the bottom panels: the Sub Red subsample for $(G-\grp)$ and the Sub Mix subsample for $(\gbp-\grp)$ and $(\gbp-G)$. The Sub Red subsample includes more stars than Sub Mix, resulting in a much higher density of stars along the red $(G-\grp>1.3)$ portion of the main sequence.}
\label{fig:CMD_after_cuts}
\end{center}
\end{figure*}

To validate all the quality cuts we defined, we plot the color--magnitude diagrams for $M_{\rm G}$ as a function of the three \textit{Gaia} colors ($G-\grp$, $\gbp-\grp$ and $\gbp-G$) with and without the previously discussed astrometric and photometric cuts in Figure \ref{fig:CMD_after_cuts}. The photometric cuts remove dramatic outliers in color and magnitude space in each color and magnitude combination, indicating that they have reliably selected good quality photometry. Due to the low quality of the $\gbp$ band for the reddest, faintest stars, there is a higher density of red $(G-\grp>1.3)$ stars in the Sub Red subsample shown on the $(G-\grp)$ diagram. Those stars fall below the main sequence for $(\gbp-\grp)$ and $(\gbp-G)$ in the color--magnitude diagrams without quality cuts applied (Top panels of Figure~\ref{fig:CMD_after_cuts}). 

\section{\textit{Gaia} DR2 colors and absolute magnitudes of M and L dwarfs}
\label{sec:mlgaiadr2_colors_absmag}

Characterizing the relationships between spectral types and both colors and absolute magnitudes facilitates the classification of new objects and aids in the identification of interesting outliers. Therefore we used the MLSDSS-GaiaDR2 sample with good photometry and astrometry described in Section \ref{sec:MLSDSS} to calculate \textit{Gaia} DR2 mean colors and absolute magnitudes for M and early-L dwarfs as a function of spectral subtype. The stellar locus in color-color space is also an important tool to classify stars and identify sources of contamination, so we examined the SDSS-2MASS-Gaia stellar locus for M and L dwarfs. Finally we plotted the color--magnitude diagram using \textit{Gaia} bands to study fundamental properties of the stars.

\subsection{Mean Colors with \textit{Gaia} DR2}
\label{subsec:colorgaia}

\begin{figure*}[ht!]
\begin{center}
\includegraphics[width=0.5\linewidth]{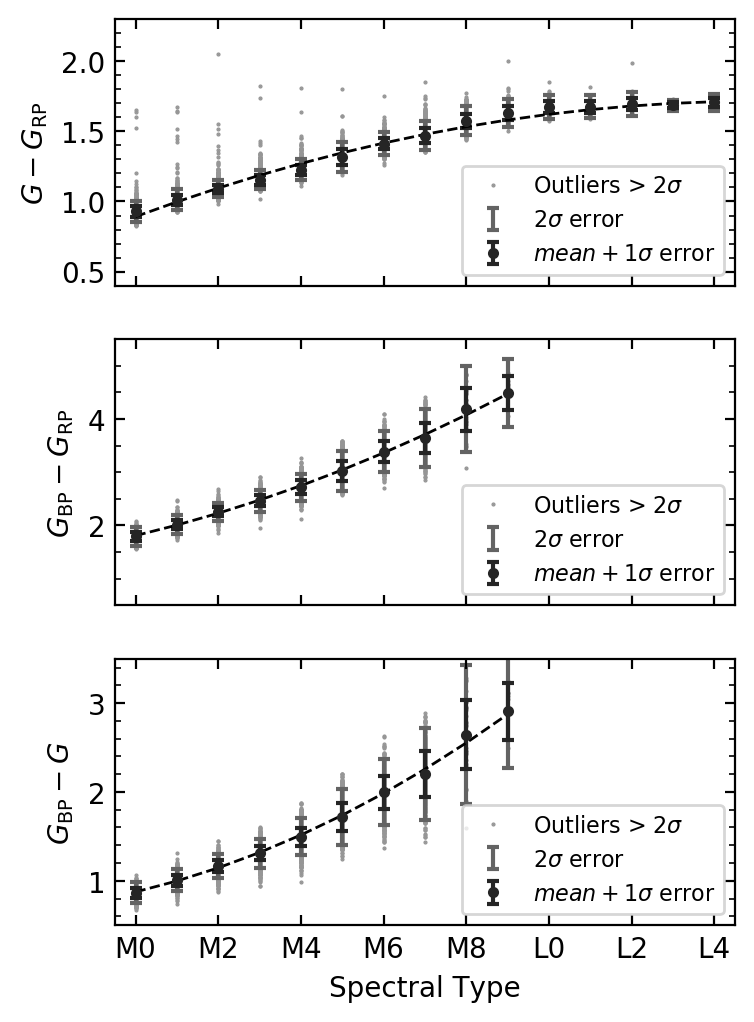}%
\includegraphics[width=0.5\linewidth]{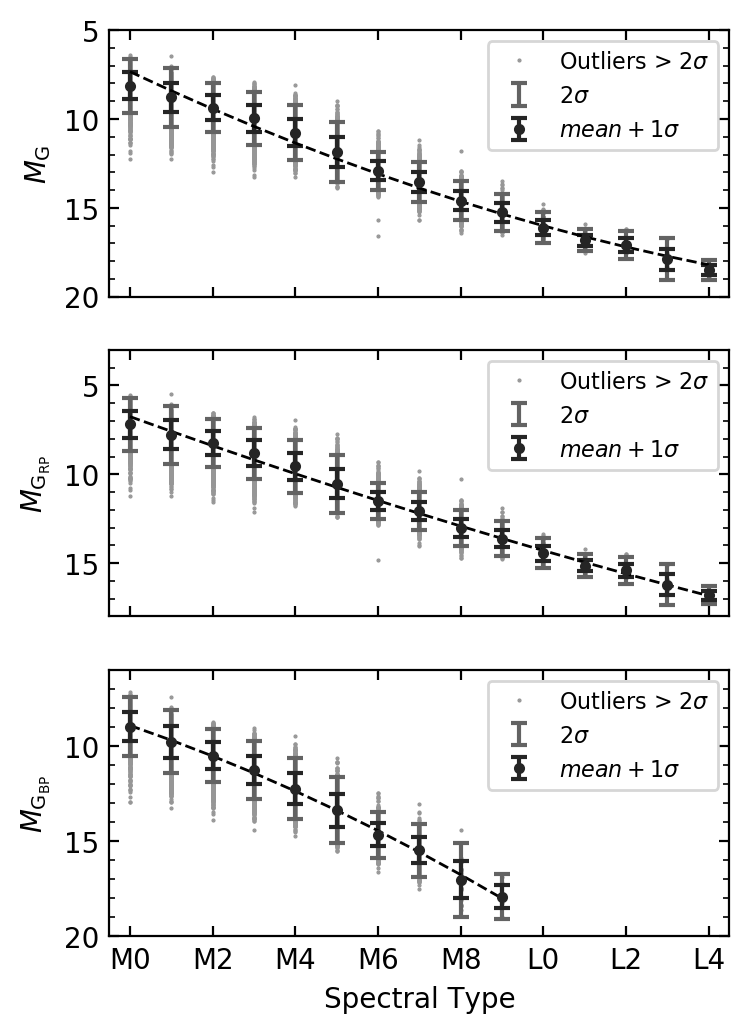}
\caption{Left panels: Distribution of \textit{Gaia} DR2 colors for the three \textit{Gaia} colors as a function of spectral type. We used the photometric subsample Sub Red for $(G- \grp)$ and the Sub Mix subsample for $(\gbp-\grp)$ and $(\gbp-G)$ (described in Section \ref{subsubsec:photcuts}). Right panels: Distribution of \textit{Gaia} DR2 Absolute Magnitudes for the three photometric \textit{Gaia}-bands $G$, $\grp$ and $\gbp$ as a function of spectral type, using the MLSDSS-GaiaDR2 sample. We used the photometric subsamples Sub G, Sub Red and Sub Mix for $M_{\rm G}$, $M_{\rm G_{RP}}$ and $M_{\rm G_{BP}}$ respectively (described in Section \ref{subsubsec:photcuts}). For all the panels we also show the mean values and $1\sigma$ and $2\sigma$ dispersion, where $\sigma$ is the standard deviation. The best fit 3rd degree polynomial to the mean values is in a black dashed line and the best fit polynomial parameters are listed in Table \ref{table:fit_abs}.}
\label{fig:mean_absmag}
\end{center}
\end{figure*}

Quantifying the correlation between the new \textit{Gaia} colors and spectral type is essential to classify new objects and detect outliers. We calculated \textit{Gaia} DR2 mean colors as a function of spectral type for $(G- \grp)$ using the Sub Red sample and $(\gbp-\grp)$ and $(\gbp-G)$ using the Sub Mix sample (See Section \ref{subsubsec:photcuts}). We removed $1,680$ stars with extinction correction $E(r-z)>0.1$ to avoid photometry contaminated by dust in front of the star.
The resulting means and standard deviations are shown in Figure~\ref{fig:mean_absmag} and enumerated in Table~\ref{table:mean_absmag}. We fit a second degree polynomial to the mean color values as a function of spectral type (shown in  Figure \ref{fig:mean_absmag}) and give the best fit parameters in Table \ref{table:fit_abs}, where $\sigma$ is the standard deviation of the stars in each bin, which was used to weight the fit. We calculated mean values for M0--L4 for the $(G - \grp)$ color, and for M0--M9 for the other two colors because L dwarfs are too faint in the $\gbp$ band and did not pass the quality cuts defined in Section \ref{subsec:qualitycuts}. 

The $(G - \grp)$ color has the tightest relation to spectral type, as shown in the top left panel in Figure~\ref{fig:mean_absmag}. The $(\gbp - \grp)$ and $(\gbp-G)$ colors have a tight relation for M0 to M7 stars, however, the dispersion increases for later spectral types. Therefore we conclude the $(G - \grp)$ color is the best proxy for spectral type for late-M and L dwarfs in the \textit{Gaia} bands.

The $(G - \grp)$ color locus in Figure \ref{fig:mean_absmag} has $36$ of its most extreme outliers redward (above) of the mean. These outliers have good photometry in $\grp$ and $G$ bands according to the quality cuts described in Section \ref{subsubsec:photcuts}, but they have low signal-to-noise fluxes in the blue band, $\gbp$ ($\texttt{mean\_flux\_over\_error}<10$). Inspection of the images of these $36$ dwarfs showed that they are binaries or have a close neighbor, which might be causing the excess in the color. These objects were not removed by the excess cut made in Section \ref{subsec:qualitycuts} because they have low SNR $\gbp$ photometry. By studying the images, we also confirmed that they were not mismatches. Furthermore, we could not find any peculiarities by plotting these objects in color-color plots for SDSS colors. We concluded the color excess is likely due to contamination in the $\grp$ band and we removed them from the analysis. In the MLSDSS-GaiaDR2 sample, these objects are indicated as $\texttt{photometric\_sample\_subg} = 2$, $\texttt{photometric\_sample\_subred} = 2$ and $\texttt{photometric\_sample\_submix} = 2$.  




\begin{longrotatetable}
\begin{deluxetable*}{c|ccc|ccc|ccc|ccc|ccc|ccc}
\tablecaption{Mean \textit{Gaia} DR2 Absolute Magnitudes and Colors for MLSDSS-GaiaDR2 \label{table:mean_absmag}}
\tablecolumns{10}
\tablehead{\multicolumn{1}{c}{SpT} &  \multicolumn{3}{c}{$G-\grp$} &  \multicolumn{3}{c}{$\gbp-\grp$} &  \multicolumn{3}{c}{$\gbp-G$} &  \multicolumn{3}{c}{$M_{\rm G}$} &  \multicolumn{3}{c}{$M_{\rm G_{RP}}$} &  \multicolumn{3}{c}{$M_{\rm G_{BP}}$} \\ \colhead{} & \colhead{N} & \colhead{mean} & \colhead{$\sigma$} & \colhead{N} & \colhead{mean} & \colhead{$\sigma$} & \colhead{N} & \colhead{mean} & \colhead{$\sigma$} & \colhead{N} & \colhead{mean} & \colhead{$\sigma$} & \colhead{N} & \colhead{mean} & \colhead{$\sigma$} & \colhead{N} & \colhead{mean} & \colhead{$\sigma$}}

\startdata
M0 & 1600 & 0.93 & 0.04 & 1599 & 1.79 & 0.09 & 1599 & 0.87 & 0.06 & 1620 & 8.13 & 0.76 & 1600 & 7.2 & 0.74 & 1599 & 8.99 & 0.78 \\
M1 & 1473 & 1.01 & 0.04 & 1468 & 2.02 & 0.09 & 1468 & 1.0 & 0.06 & 1506 & 8.78 & 0.83 & 1473 & 7.76 & 0.82 & 1468 & 9.78 & 0.84 \\
M2 & 2937 & 1.09 & 0.03 & 2934 & 2.25 & 0.09 & 2934 & 1.16 & 0.07 & 2989 & 9.35 & 0.69 & 2937 & 8.26 & 0.67 & 2934 & 10.51 & 0.7 \\
M3 & 3651 & 1.16 & 0.03 & 3624 & 2.46 & 0.1 & 3624 & 1.31 & 0.08 & 3698 & 9.97 & 0.74 & 3651 & 8.81 & 0.72 & 3624 & 11.27 & 0.75 \\
M4 & 2855 & 1.23 & 0.04 & 2790 & 2.72 & 0.13 & 2790 & 1.49 & 0.1 & 2889 & 10.77 & 0.77 & 2855 & 9.55 & 0.75 & 2790 & 12.25 & 0.8 \\
M5 & 1400 & 1.32 & 0.05 & 1131 & 3.02 & 0.19 & 1131 & 1.72 & 0.16 & 1416 & 11.86 & 0.85 & 1400 & 10.54 & 0.81 & 1131 & 13.38 & 0.87 \\
M6 & 2488 & 1.41 & 0.04 & 969 & 3.39 & 0.19 & 969 & 2.0 & 0.19 & 2510 & 12.92 & 0.53 & 2488 & 11.5 & 0.5 & 969 & 14.67 & 0.61 \\
M7 & 2645 & 1.47 & 0.05 & 527 & 3.64 & 0.27 & 527 & 2.2 & 0.26 & 2686 & 13.54 & 0.56 & 2645 & 12.07 & 0.52 & 527 & 15.48 & 0.7 \\
M8 & 1093 & 1.57 & 0.05 & 33 & 4.18 & 0.41 & 33 & 2.65 & 0.39 & 1122 & 14.6 & 0.55 & 1093 & 13.02 & 0.51 & 33 & 17.04 & 0.97 \\
M9 & 354 & 1.63 & 0.05 & 8 & 4.48 & 0.32 & 8 & 2.91 & 0.32 & 363 & 15.26 & 0.53 & 354 & 13.62 & 0.5 & 8 & 17.92 & 0.6 \\
L0 & 119 & 1.67 & 0.04 & 0 & $\cdots$ & $\cdots$ & 0 & $\cdots$ & $\cdots$ & 121 & 16.11 & 0.44 & 119 & 14.45 & 0.42 & 0 & $\cdots$ & $\cdots$ \\
L1 & 46 & 1.68 & 0.04 & 0 & $\cdots$ & $\cdots$ & 0 & $\cdots$ & $\cdots$ & 47 & 16.82 & 0.31 & 46 & 15.14 & 0.31 & 0 & $\cdots$ & $\cdots$ \\
L2 & 16 & 1.7 & 0.04 & 0 & $\cdots$ & $\cdots$ & 0 & $\cdots$ & $\cdots$ & 16 & 17.11 & 0.4 & 16 & 15.42 & 0.39 & 0 & $\cdots$ & $\cdots$ \\
L3 & 6 & 1.69 & 0.02 & 0 & $\cdots$ & $\cdots$ & 0 & $\cdots$ & $\cdots$ & 6 & 17.89 & 0.59 & 6 & 16.21 & 0.58 & 0 & $\cdots$ & $\cdots$ \\
L4 & 3 & 1.71 & 0.03 & 0 & $\cdots$ & $\cdots$ & 0 & $\cdots$ & $\cdots$ & 3 & 18.51 & 0.27 & 3 & 16.8 & 0.25 & 0 & $\cdots$ & $\cdots$ \\
L5 & 0 & $\cdots$ & $\cdots$ & 0 & $\cdots$ & $\cdots$ & 0 & $\cdots$ & $\cdots$ & 0 & $\cdots$ & $\cdots$ & 0 & $\cdots$ & $\cdots$ & 0 & $\cdots$ & $\cdots$ \\
L6 & 1 & 1.77 & 0.0 & 0 & $\cdots$ & $\cdots$ & 0 & $\cdots$ & $\cdots$ & 2 & 18.92 & 0.17 & 1 & 16.98 & 0.0 & 0 & $\cdots$ & $\cdots$ \\
L7 & 0 & $\cdots$ & $\cdots$ & 0 & $\cdots$ & $\cdots$ & 0 & $\cdots$ & $\cdots$ & 0 & $\cdots$ & $\cdots$ & 0 & $\cdots$ & $\cdots$ & 0 & $\cdots$ & $\cdots$ \\
\enddata
\tablecomments{Number of objects included in calculation (N), mean color or magnitude, and standard deviation ($\sigma$) of the mean. }
\end{deluxetable*}
\end{longrotatetable}

\begin{deluxetable*}{cccccc}[ht!]
\tablecaption{Best fit parameters for \textit{Gaia} magnitudes and colors as a function of spectral type \label{table:fit_abs}}
\tablehead{\colhead{Band/Color} & \colhead{$a$} & \colhead{$b$} & \colhead{$c$} & \colhead{$\sigma$} & \colhead{Valid Range}}
\startdata
$G - \grp$ & $-0.0036\pm 0.0005$ & $0.11\pm 0.01$ & $0.89\pm 0.02$ & 0.03 & M0$<$SpT$<$L4 \\
$\gbp - \grp$ & $0.012\pm 0.002$ & $0.19\pm 0.01$ & $1.81\pm 0.02$ & 0.09 & M0$<$SpT$<$M9 \\
$\gbp - G$ & $0.012\pm 0.001$ & $0.11\pm 0.01$ & $0.87\pm 0.01$ & 0.08 & M0$<$SpT$<$M9 \\
$M_G$ & $-0.023\pm 0.003$ & $1.1\pm 0.1$ & $7.3\pm 0.3$ & 0.52 & M0$<$SpT$<$L4 \\
$M_{\grp}$ & $-0.008\pm 0.003$ & $0.8\pm 0.1$ & $6.8\pm 0.2$ & 0.47 & M0$<$SpT$<$L4 \\
$M_{\gbp}$ & $0.03\pm 0.01$ & $0.7\pm 0.1$ & $8.9\pm 0.1$ & 1.24 & M0$<$SpT$<$M9 \\
\enddata
\tablecomments{Results from the best fit to the mean absolute magnitudes and colors as a quadratic function of spectral type, $a\times {\rm SpT}^2 + b\times {\rm SpT}+ c$, with M0=0, M9=9, and L4=14, as shown in Figure \ref{fig:mean_absmag} }
\end{deluxetable*}

\subsection{Mean Absolute Magnitudes with \textit{Gaia} DR2}
\label{subsec:absmaggaia}

We used the MLSDSS-GaiaDR2 sample to calculate mean absolute magnitudes in the three \textit{Gaia} DR2 bands as a function of spectral subtype of M and L dwarfs using \textit{Gaia} DR2 parallaxes. We chose the appropriate photometric subsample described in Section \ref{subsubsec:photcuts} for each band: Sub G for $M_{\rm G}$, Sub Red for $M_{\rm G_{RP}}$ and Sub Mix for $M_{\rm G_{BP}}$. As in the previous section, we removed stars with high extinction corrections ($E(r-z)>0.1$) to minimize photometry contaminated by foreground dust.
The distributions of absolute magnitudes as a function of spectral subtype are shown in Figure~\ref{fig:mean_absmag} right panels and listed in Table~\ref{table:mean_absmag}. 

Note that with the parallax SNR cut applied to the astrometric subsample, we selected a maximum of $10 \%$ uncertainty in distance, which corresponds to a maximum of $0.2$~mag uncertainty in absolute magnitude ($\texttt{parallax\_over\_error} > 10$, see Section \ref{subsec:qualitycuts} for details, \citealt{Lindegren2018}).

For the $\gbp$~band, the standard deviation of the distribution of absolute magnitudes per spectral type ($\sigma$) increases towards later spectral types as shown in Table \ref{table:mean_absmag}. This effect is due to the higher uncertainties in the $\gbp$ band for fainter, redder stars (M0 stars have a mean flux SNR in the blue band of $\texttt{phot\_bp\_mean\_flux\_over\_error} = 137$ and M8 of $\texttt{phot\_bp\_mean\_flux\_over\_error} = 12$). In the $\grp$ and $G$~bands, the standard deviation for late M dwarfs is one order of magnitude smaller than in the $\gbp$~band. This is because the photometry in $\grp$ and $G$~bands have higher SNR than the blue band for our sample (M0 stars have a mean flux SNR in the red band of $\texttt{phot\_rp\_mean\_flux\_over\_error} = 353$ and in the $G$~band of $\texttt{phot\_g\_mean\_flux\_over\_error} = 1274$ and M8 $\texttt{phot\_rp\_mean\_flux\_over\_error} = 71$ and $\texttt{phot\_g\_mean\_flux\_over\_error} = 279$, respectively). 

To quantify the relationship between absolute magnitude and spectral type, we fit a second degree polynomial to the mean values as a function of spectral type, shown as a black dashed line in Figure \ref{fig:mean_absmag}. We used the $\sigma$ as a weight to perform the fit. The best fit parameters for the polynomial are given in Table \ref{table:fit_abs}. 

While the vast majority of the sample is well-characterized by a second degree polynomial in absolute magnitude versus spectral type, there are $\sim 1000$ outliers ($\sim 4.7 \%$) more than $2\sigma$ away from the mean in each plot in the right panel of Figure \ref{fig:mean_absmag}. For spectral type earlier than M3, the majority of outliers are fainter than the average fit, and the opposite is the case after M3. We surmise that the scatter towards fainter absolute magnitudes for earlier spectral types is associated with low metallicity, halo or thick disk stars, while the scatter towards brighter absolute magnitudes is related to high metallicity, magnetic activity and/or unresolved binarity. We discuss the relation to age of these particular features in Section \ref{sec:age}.

\subsection{The SDSS-2MASS-\textit{Gaia} M and L Dwarf Stellar Locus in Color Space}

\begin{figure*}[ht!]
\begin{center}
\includegraphics[width=\linewidth]{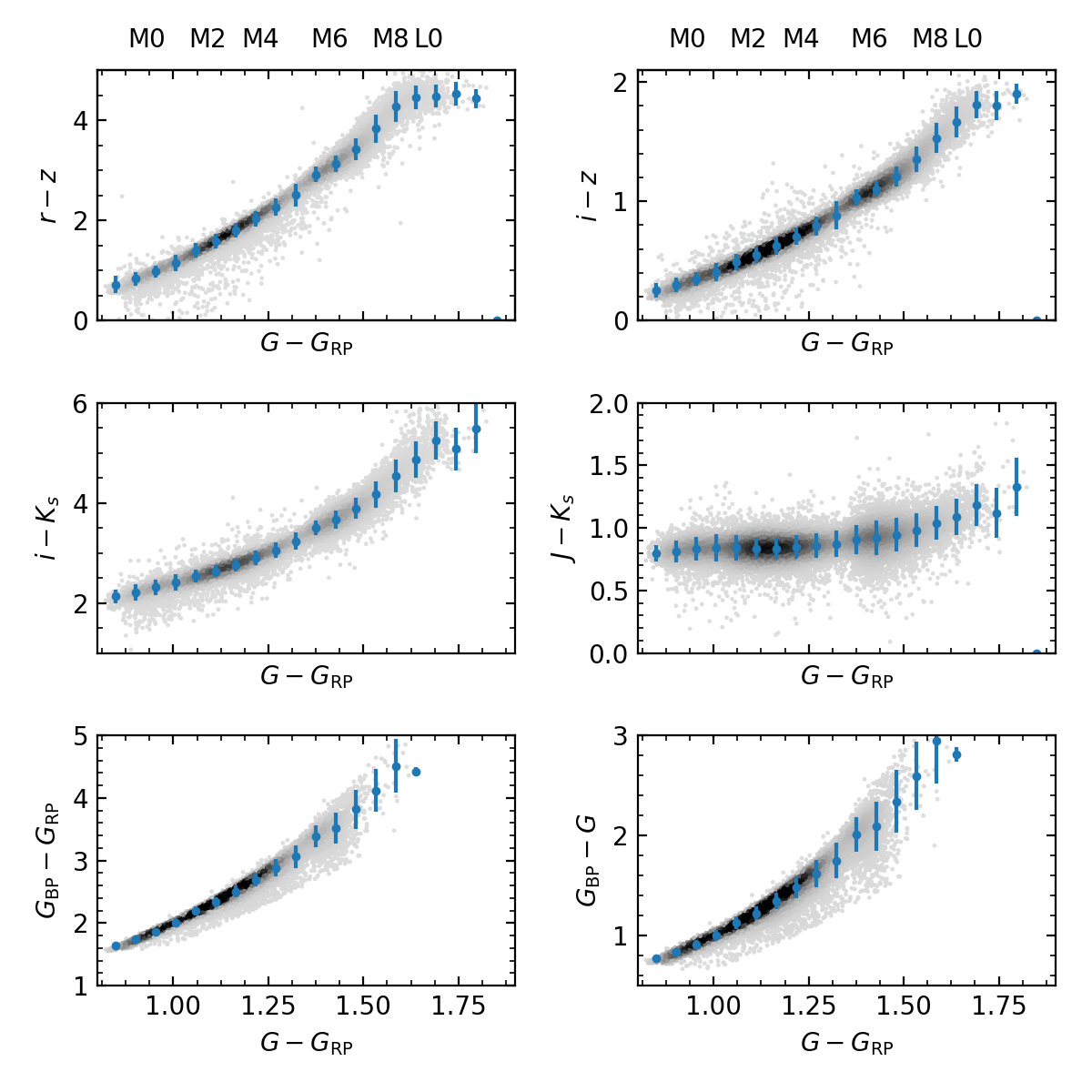}
\caption{SDSS-2MASS-\textit{Gaia} M and L dwarfs color-color stellar locus for $(r-z)$, $(i-z)$, $(i-K_s)$ and $(J-K_s)$ for individual stars in density contours (grey) and for mean and standard deviation of the locus (blue points with error bars). We used $(G-\grp)$ to calculate the locus, so we chose the photometric subsample Sub Red discussed in Section \ref{subsubsec:photcuts} and selected stars with good SDSS photometry.}
\label{fig:locus}
\end{center}
\end{figure*}

Previous work \citep[e.g.][]{Covey2007,Davenport2014} has shown the power of characterizing the color-color space of the stellar locus to classify stars, detect sources of contamination in a sample, and to calculate extinction corrections. A characterized stellar locus for \textit{Gaia} colors of M and L dwarfs provides a continuous parametrization of color as a function of effective temperature and facilitates finding color outliers for follow up. Furthermore, incorporating photometry from other surveys provides a relation between colors that will allow us to estimate \textit{Gaia} DR2 photometry for M or L dwarfs from other catalogs colors, or vice versa. We used the MLSDSS-GaiaDR2 sample which contains \textit{Gaia} ($G$, $\gbp$ and $\grp$), SDSS ($u$, $g$, $r$, $i$ and $z$) and 2MASS ($J$, $H$ and  $K_s$) photometry to search for an optimal characterization locus for M and L dwarfs. The characterized stellar locus is shown for $(r-z)$, $(i-z)$, $(i-K_s)$, $(J-K_s)$, $(\gbp-\grp)$ and $(\gbp-G)$ in Figure \ref{fig:locus}.

We chose $(G-\grp)$ as a grounding color because it has the tightest relation to spectral type (see Section \ref{subsec:colorgaia}). We used the appropriate photometric subsample described in Section \ref{subsubsec:photcuts} for each plot: Sub Red for $(r-z)$, $(i-z)$, $(i-K_s)$, $(J-K_s)$ and Sub Mix for $(\gbp-G)$ and $(\gbp-\grp)$. We also selected stars with good SDSS photometry using the $\texttt{GOODPHOT\_SDSS}$ flag (see Section \ref{subsec:basemlsdss} for more details on this cut), including the highest possible number of objects with good photometry in the analysis (median SNR $\sim 900$ for SDSS photometry, $\sim 300$ for 2MASS photometry, $\sim 700$ for $G$~band, $\sim 200$ for the $\grp$~band, $\sim 50$ for the $\gbp$~band). We modeled the sequence using a step of $\delta (G-\grp) = 0.05$ for the full color range covered by M and L dwarfs ($0.8<(G-\grp)<2.0$). 

Most of the colors have a linear, non-zero-slope relation with $(G-\grp)$. The $(r-z)$ and $(i-z)$ linear relations are consistent with previous work \citep{Covey2007,Davenport2014,Schmidt2015}. $(J-K_s)$ has a flat relation with $(G-\grp)$ for early M and a slightly positive slope for $(G-\grp)>1.4$, which indicates it is not a good color to distinguish spectral type. This result is consistent with the conclusions in \cite{Schmidt2015} for $(J-H)$. The linear relation between $(r-z)$, $(i-z)$, $(i-K_s)$ and $(G-\grp)$ breaks for L dwarfs  at $(G-\grp)\sim 1.7$. We will discuss this break in Section \ref{sec:mlinsdss2mass}. Finally, $(\gbp-\grp)$ and $(\gbp-G)$ have a tight linear relation with $(G-\grp$). The low dispersion of outliers for these colors is due to the photometric cuts applied to create the subsample Sub Mix.  

\subsection{Gaia DR2 Color Magnitude diagrams}
\label{subsec:cmdmlsdssgaiadr2}

\begin{figure*}[ht!]
\begin{center}
\includegraphics[width=\linewidth]{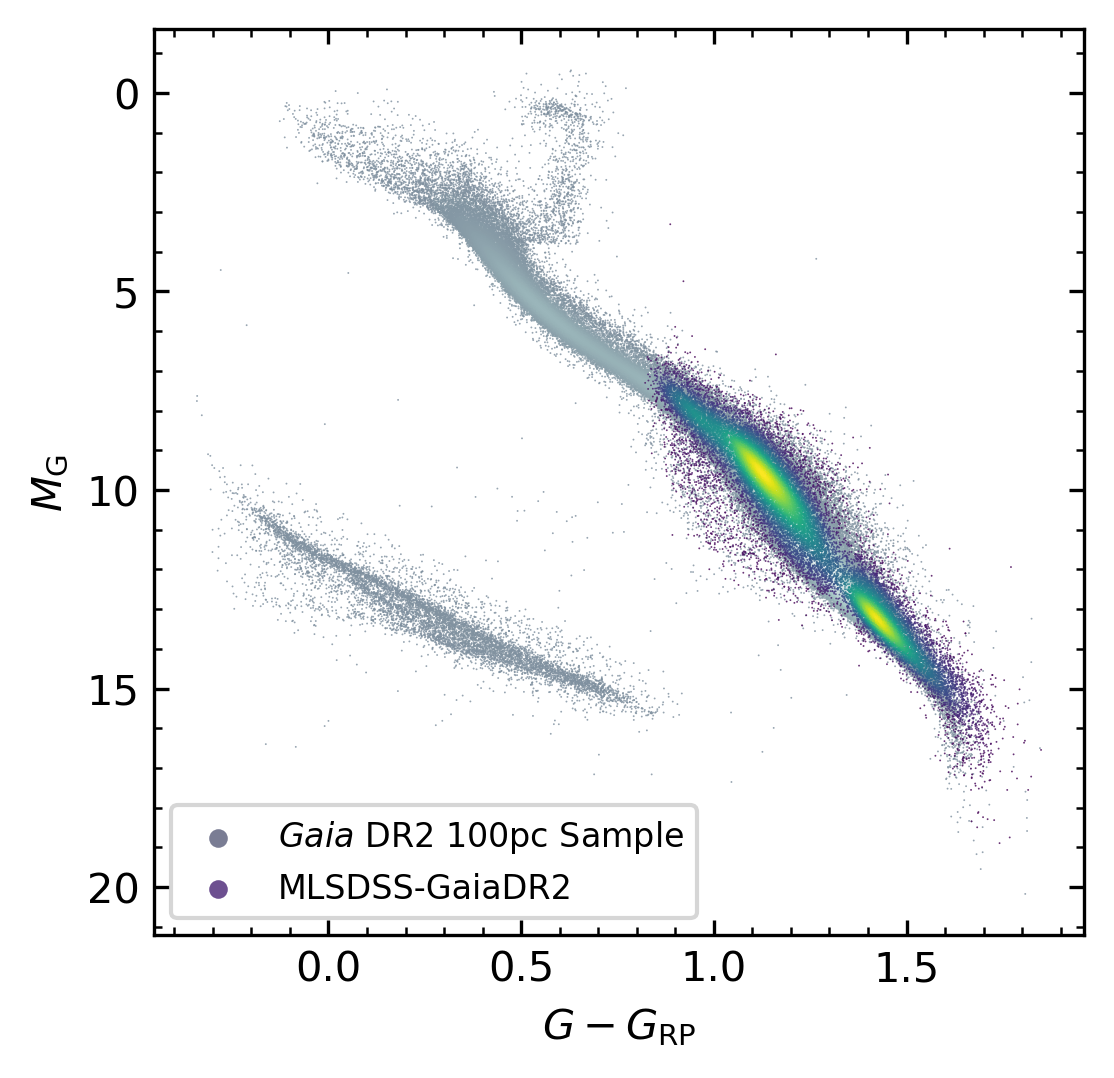}
\caption{Color magnitude diagram for the MLSDSS-GaiaDR2 sample color-coded with its density map with the quality cuts described in \ref{subsec:qualitycuts} applied, compared to $100$pc Sample from \textit{Gaia} DR2 in gray, also color-codded with its density map. To clean the MLSDSS-GaiaDR2 sample we used the quality cuts in the photometric subsample Sub Red.}
\label{fig:colormag}
\end{center}
\end{figure*}

To put the MLSDSS-GaiaDR2 sample in broader stellar context, we compare it to the solar neighborhood ($\le 100$ pc) sample in the $M_{\rm G}$ vs $(G-\grp)$ color--magnitude diagram (CMD, Figure \ref{fig:colormag}). 
Given that we used the red \textit{Gaia} color $(G-\grp)$, we used the photometric subsample Sub Red to use the highest number of stars with good photometry in the analysis. We chose this color because it has the tightest relation with spectral type (see Figure \ref{fig:mean_absmag}) and we decided to use $M_{\rm G}$ because this band has the smallest photometric error for all stars in the MLSDSS-GaiaDR2 sample (see Figure \ref{fig:flux_over_error}). 
The solar neighborhood sample shows the full main sequence as well as the beginning of the red giant branch and the white dwarf sequence, while M and L stars from the MLSDSS-GaiaDR2 sample fall at the faint, red end of the main sequence. 

\begin{figure*}[ht!]
\begin{center}
\includegraphics[width=\linewidth]{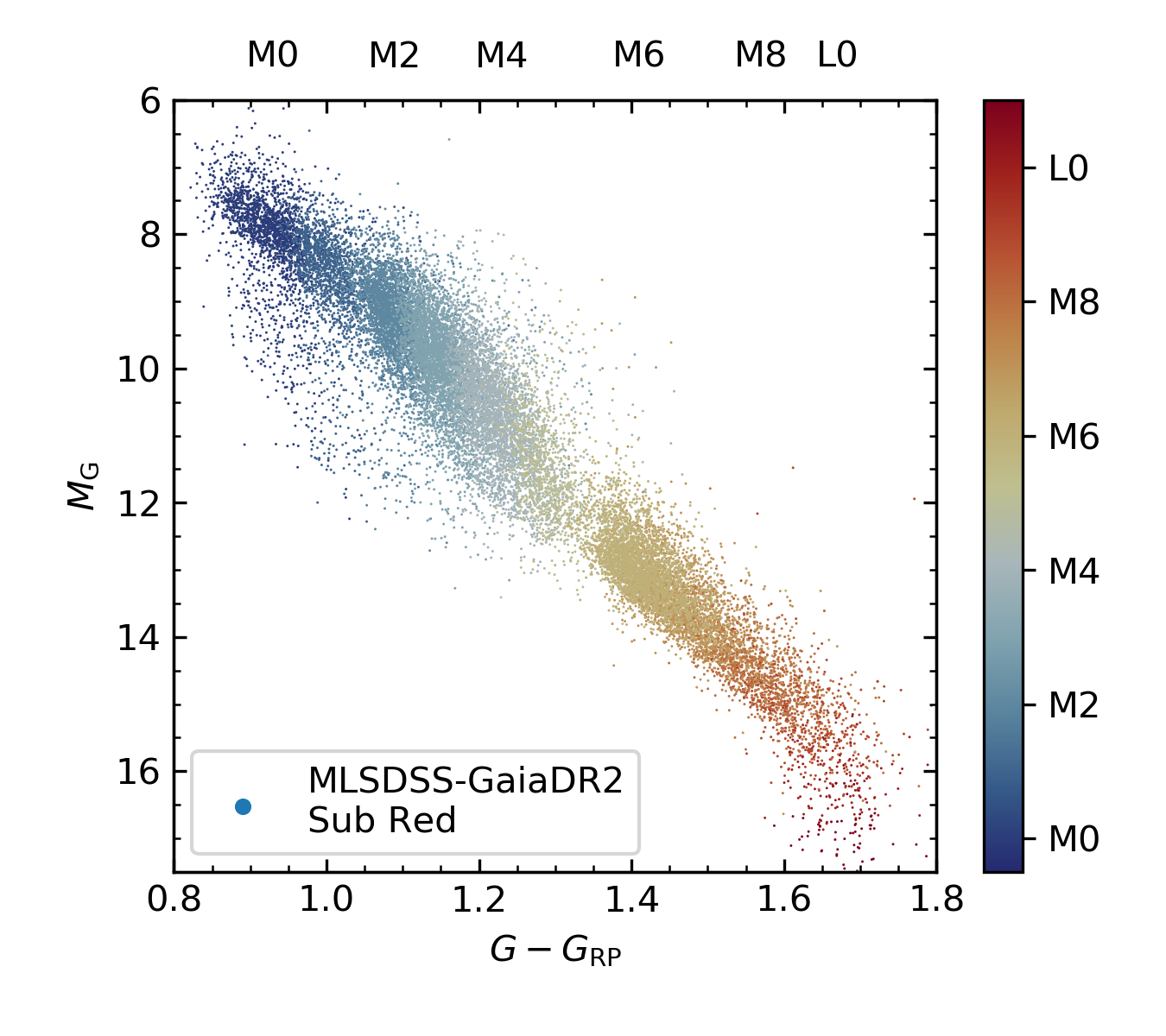}
\caption{Color-magnitude diagram for the MLSDSS-GaiaDR2 sample, color-coded by spectroscopic spectral type. To clean the MLSDSS-GaiaDR2 sample we used the quality cuts in the photometric subsample Sub Red, as in for Figure~\ref{fig:colormag}.}
\label{fig:colormag2}
\end{center}
\end{figure*}

The MLSDSS-GaiaDR2 sample color--magnitude diagram is shown again in Figure \ref{fig:colormag2} with SDSS spectral type color coded. Note that the gap around M5 is due to SDSS selection effects which were pointed out by \citet{West2011}. In this figure, the main sequence widens for stars fainter than $(G-\grp) \sim 1.05$.
This effect could be related to the transition into fully convective
low mass stars. 

\begin{figure}[ht!]
\begin{center}
\includegraphics[width=\linewidth]{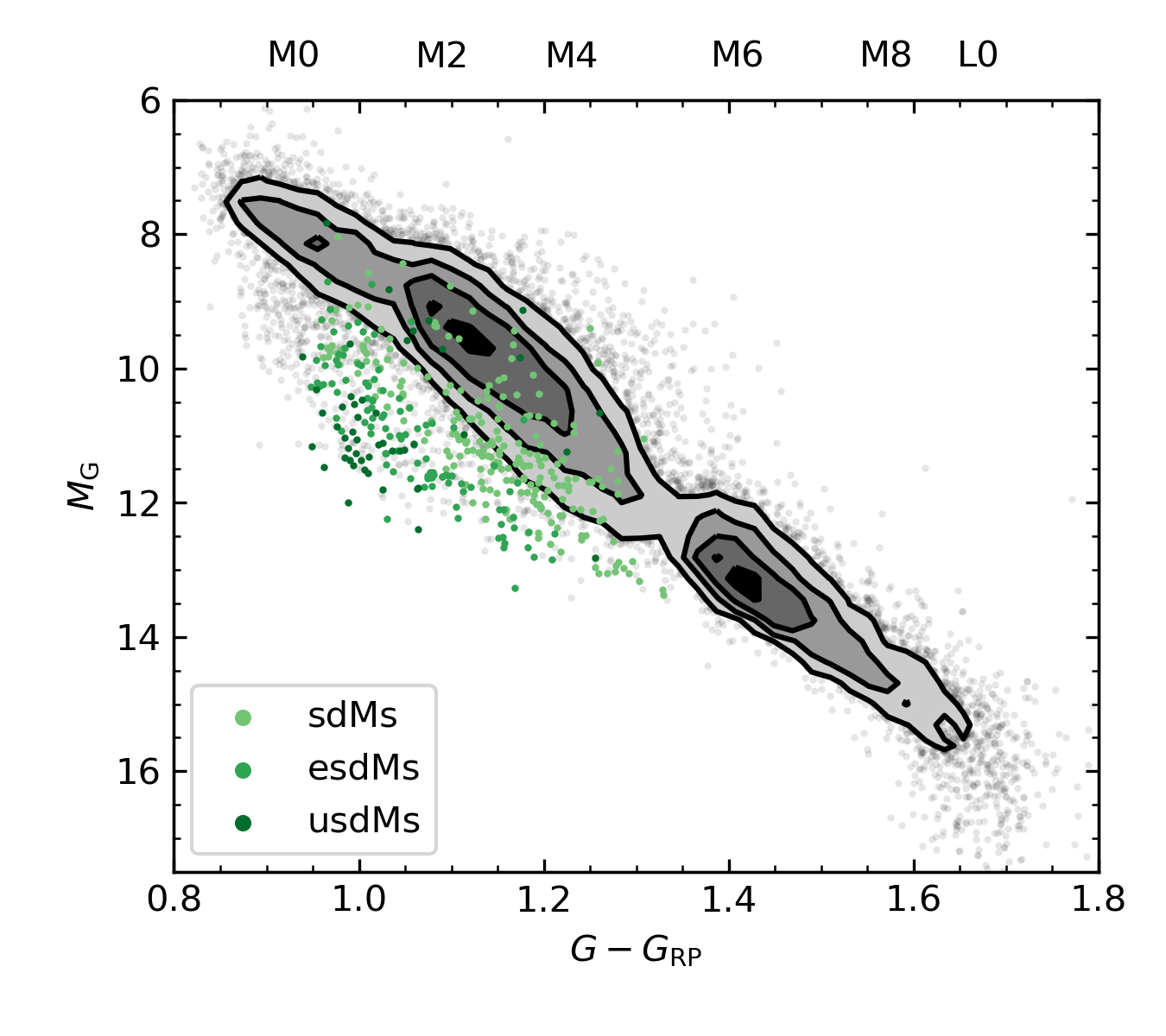}
\caption{Contours in gray of the color-magnitude diagram of MLSDSS-GaiaDR2 with the $376$ matches with the subdwarfs from \citet{Savcheva2014}. We distinguish between subdwarfs (sdMs; light green), extreme subdwarfs (esdMs; green) and ultrasubdwarf (usdMs; dark green) as assigned by \citet{Savcheva2014} using the metallicity proxy $\zeta$. We note that most subdwarfs fall below the main sequence as expected.}
\label{fig:colormagwsubdwarfs}
\end{center}
\end{figure}

For earlier-type stars ($G-\grp<1.3$), there is a significant number of objects below the main sequence. These objects are likely to be low metallicity, old stars. This faint outlier population is not present for later types ($G-\grp>1.3$). This is not likely to be a result of the quality cuts we made in Section \ref{subsec:qualitycuts} because the scatter is not present when there are no cuts applied (see Figure \ref{fig:CMD_after_cuts}). The lack of scatter in the region where subdwarfs typically lie for later types could be a selection effect from SDSS or a physical difference in the colors of later-type subdwarfs \citep[which could be less blue than their earlier spectral type counterparts][]{West2004,Lepine2008}. 
To corroborate that the stars scattered below the main sequence are primarily subdwarfs, we cross-matched MLSDSS-GaiaDR2 with the catalog of subdwarfs from \citet{Savcheva2014}. 
From this cross-match we found $376$ subdwarfs in the MLSDSS-GaiaDR2 sample which had good photometry and astrometry according to our quality cuts (Section \ref{subsec:qualitycuts}). In Figure~\ref{fig:colormagwsubdwarfs} we show a color--magnitude diagram of these $376$ subdwarfs compared to the MLSDSS-GaiaDR2 sample. 
We also included the distinction between subdwarfs (sdMs), extreme subdwarfs (esdMs) and ultrasubdwarfs (usdMs) according the metallicity proxy ($\zeta$) \citep{Lepine2007,Dhital2012}. 
As expected, the esdMs fall the furthest below the main sequence, with the usdMs and sdMs falling progressively closer to main sequence stars, consistent with \citet{Savcheva2014}. However, some of the \citet{Savcheva2014} subdwarfs fall on the main sequence of our color--magnitude diagram. Many of these stars have relatively low signal-to-noise spectra and may have been mis-classified as subdwarfs in that work.

\citet{Bochanski2013} found that the separation between subdwarf
types was $\sim 1$ mag in $M_{\rm r}$ at a given $r-z$ color or spectral type, which is approximately the same separation we 
observe in Figure \ref{fig:colormagwsubdwarfs} for $M_{\rm G}$ at a given $G-\grp$. The position of the subdwarfs in the color--magnitude diagram is also consistent with other work on metal-poor M dwarfs (see for e.g. 
\citealt[][]{Lepine2007,Jao2008,Jao2017}).

There are also sources scattered above the main sequence, likely caused by M dwarf binaries, high metallicity, magnetic activity (see Section~\ref{subsection:halphcmd}), and dust contamination in \textit{Gaia} bands (see Section~\ref{subsec:crossmatch}). Note that some of the most distant stars from the main sequence could be binaries with an M dwarf primary and a giant companion \citep{Collaboration2018a}. The scatter above and below the main sequence is discussed in more detail in Section \ref{sec:age}. 

\section{M and L dwarf Absolute Magnitudes in SDSS and 2MASS Bands}
\label{sec:mlinsdss2mass}

\textit{Gaia} DR2 distances are an order of magnitude more precise than the photometric distances in MLSDSS (uncertainties of $\sim5\%$ versus $\sim20\%$), allowing us to calculate absolute magnitudes with a median error of $0.15$ mag for SDSS photometry and $0.12$ mag for 2MASS photometry. We re-calculated the relationship between absolute magnitudes and spectral type for SDSS and 2MASS photometry with these new values and the relationship between absolute magnitudes and the $(r-z)$ color. These relations are useful to estimate spectrophotometric and photometric distances for stars that are not in \textit{Gaia} DR2, as shown by previous work, which was based on less than a hundred stars with parallaxes \citep[e.g.][]{Hawley2002,Bochanski2010}. We caution, however, that because the MLSDSS-GaiaDR2 sample has not been vetted for binaries and low metallicity stars, it may be subject to biases not present in the previous, smaller samples. 

\begin{figure*}[ht!]
\begin{center}
\includegraphics[width=\linewidth]{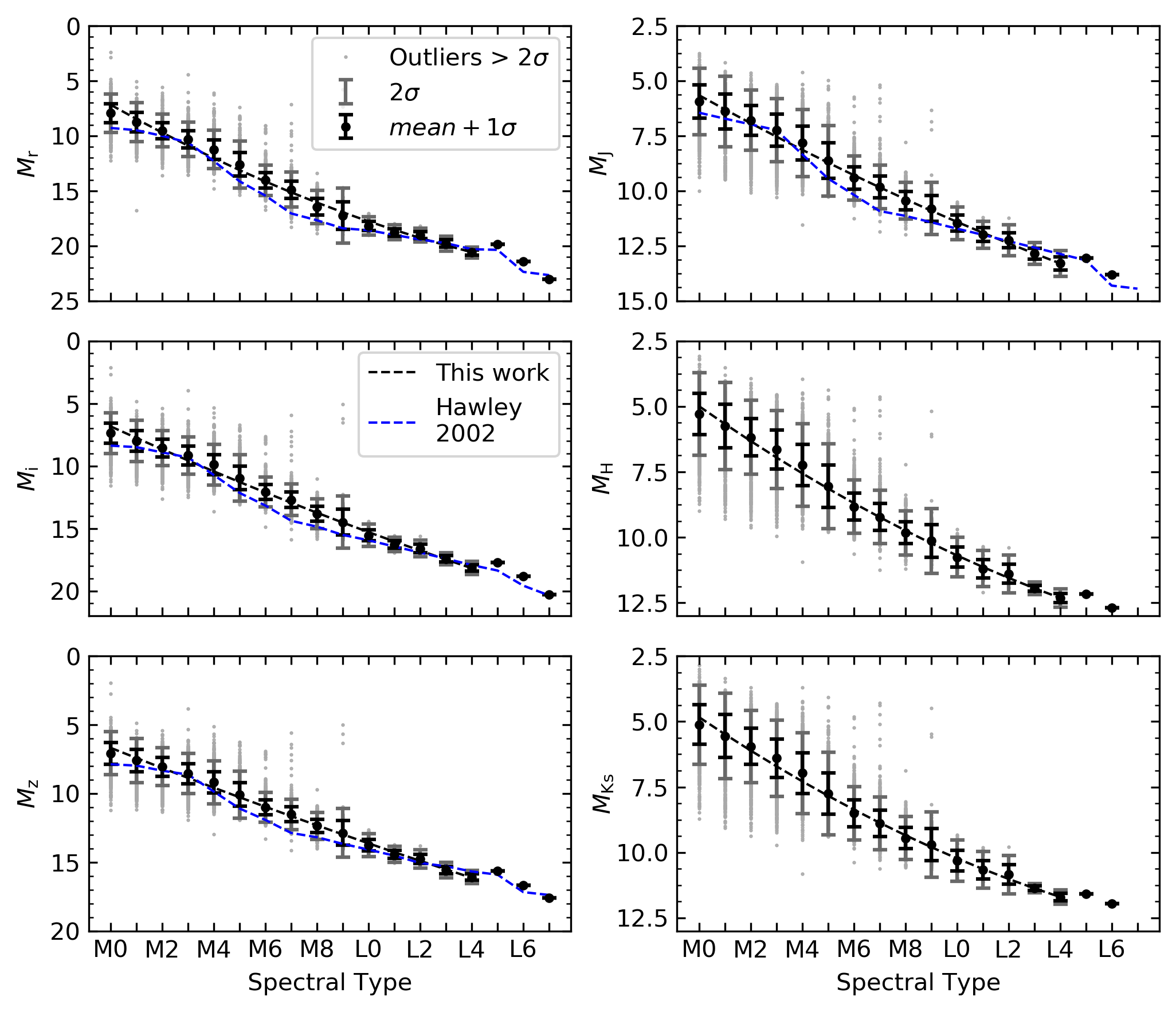}
\caption{Distribution of absolute magnitudes as a function of spectral type for SDSS and 2MASS photometry. We show the mean values and the $1\sigma$ and $2\sigma$ dispersion, where $\sigma$ is the standard deviation per bin of spectral type. In a black dashed line we show the second-order polynomial fit to these data, with parameters listed in Table \ref{table:fit_abs_mag_sdss_2mass}. In light gray we show the outliers for each relation. We include results from \citet{Hawley2002} in blue for comparison.}
\label{fig:sdss_abs_2mass_mag}
\end{center}
\end{figure*}

\begin{figure*}[ht!]
\begin{center}
\includegraphics[width=\linewidth]{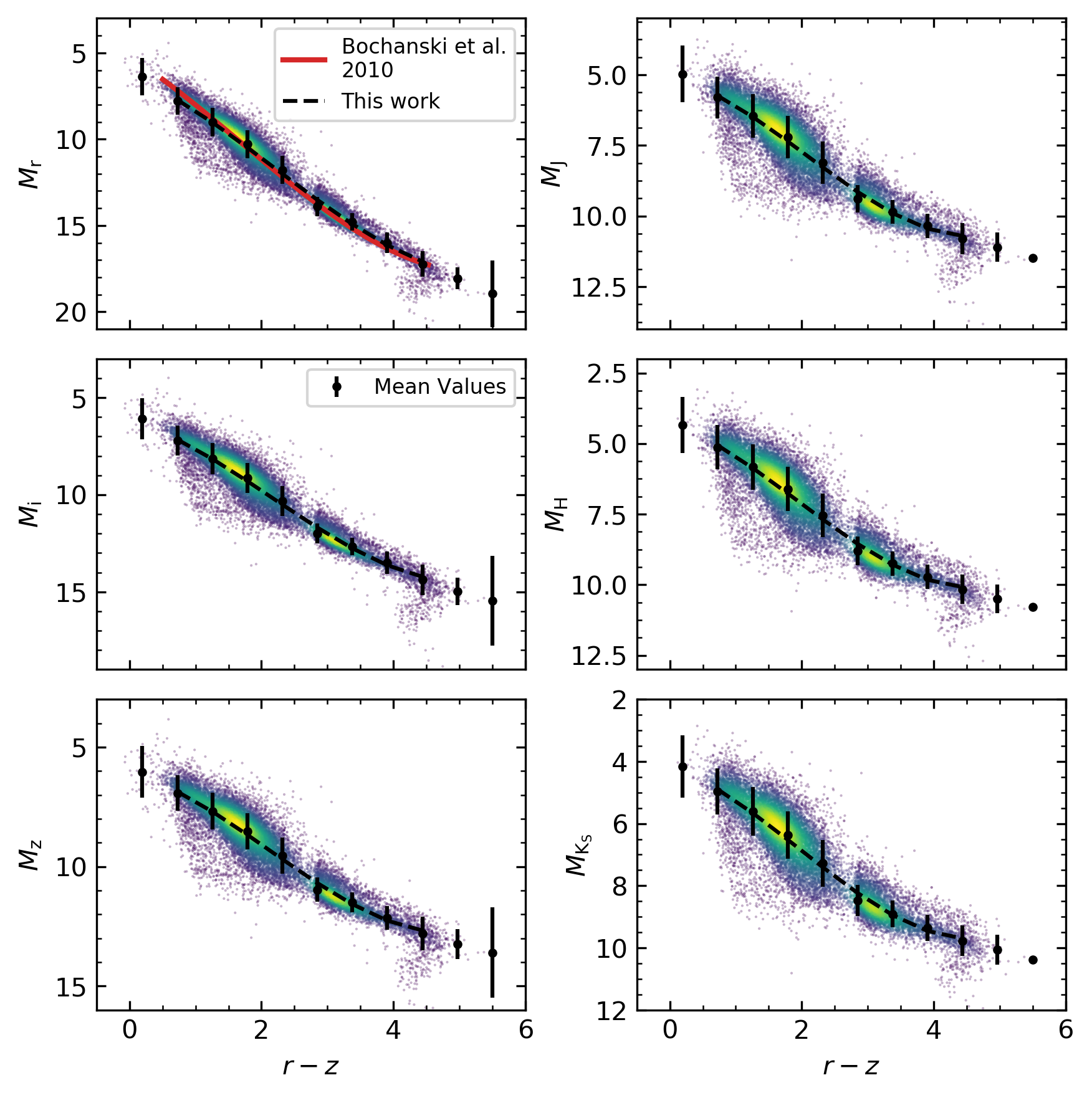}
\caption{Distribution of absolute magnitudes as a function of color for SDSS and 2MASS photometry. We show the density map of stars in gray and the mean values and standard deviation per bin of color in black. We show a third-order polynomial fit to stars between $0.5<r-z<4.5$, with parameters listed in Table \ref{table:fit_abs_mag_color_sdss}. The \citet{Bochanski2010} fit to $M_r$ as a function of $r-z$ is shown in the upper left panel, and is consistent with our fit.}
\label{fig:abs_mag_color_sdss}
\end{center}
\end{figure*}

To generate the most accurate relationship between absolute magnitudes and spectral type and color, we applied the astrometric cuts discussed in Section~\ref{subsubsec:astrometriccuts} and the photometric cut for SDSS photometry discussed in Section~\ref{subsec:basemlsdss}, thereby selecting objects with the best astrometry and photometry available. The distribution of absolute magnitudes as a function of spectral type is shown in Figure~\ref{fig:sdss_abs_2mass_mag} and as a function of color in Figure~\ref{fig:abs_mag_color_sdss}. 

For each spectral type we calculated the mean value and the standard deviation $(\sigma)$ in absolute magnitude. We also performed a fit to the mean values as a function of spectral type with $\sigma$ as weights. The fit only extends to spectral type L4 due to the small number of later-type objects; most objects later than L4 are too faint to be in \textit{Gaia} DR2 and so cannot be included (See Section~\ref{subsec:crossmatch}). The best fit parameters for SDSS $riz$ and 2MASS $JHK_s$ absolute magnitudes are listed in Table~\ref{table:fit_abs_mag_sdss_2mass}. For comparison, we included the mean values as a function of spectral type calculated by \citet{Hawley2002} from a sample of $718$ M and L dwarfs with photometric distances. We note that our fit for $M_{\rm r}$, $M_{\rm i}$ and $M_{\rm z}$ lies above the values calculated by that work. This is likely in part due to the uncertainties in the photometric distance, but also may be due to the binary population in our sample when performing the fit: binary systems with two equal mass components fall $0.7$ mag above the main sequence, which could result in brighter mean absolute magnitude.

While most of the MLSDSS-GaiaDR2 sample follows the mean trend for absolute magnitude as a function of spectral type, there are outliers in each spectral type bin that are more than $2\sigma$ from the mean in absolute magnitude. These outliers have a distribution similar to that in the \textit{Gaia} photometric color--magnitude diagram (Figure~\ref{fig:colormag2}). Those scattered to fainter absolute magnitudes can be associated with low metallicity stars and are only present for earlier spectral types. The scatter towards brighter absolute magnitudes is mostly present for early and mid-M dwarfs and can be associated with binarity, high metallicity, and/or magnetic activity. We will discuss more this scatter in Section \ref{sec:age}.

\begin{deluxetable*}{ccccc}[ht!]
\tablecaption{Best fit parameters for SDSS and 2MASS magnitudes as a function of spectral type. \label{table:fit_abs_mag_sdss_2mass}}
\tablehead{\colhead{Band} & \colhead{$a$} & \colhead{$b$} & \colhead{$c$} & \colhead{$\sigma$}}
\startdata
$M_{\rm r}$ & $-0.03\pm 0.01$ & $1.3\pm 0.1$ & $7.2\pm 0.4$ & $0.62$\\ 
$M_{\rm i}$ & $-0.009\pm 0.004$ & $0.9\pm 0.1$ & $6.8\pm 0.2$ & $0.55$\\ 
$M_{\rm z}$ & $-0.005\pm 0.003$ & $0.74\pm 0.05$ & $6.7\pm 0.2$ & $0.5$\\ 
$M_{\rm J}$ & $-0.007\pm 0.002$ & $0.65\pm 0.04$ & $5.7\pm 0.1$ & $0.47$\\ 
$M_{\rm H}$ & $-0.012\pm 0.002$ & $0.69\pm 0.03$ & $5.0\pm 0.1$ & $0.48$\\ 
$M_{\rm K_s}$ & $-0.012\pm 0.002$ & $0.66\pm 0.03$ & $4.8\pm 0.1$ & $0.47$\ 
\enddata
\tablecomments{Best fit parameters to a quadratic fit to SDSS and 2MASS absolute magnitudes as a function of spectral type, $M = a\times {\rm SpT}^2 + b\times {\rm SpT}+ c$, as shown in Figure~\ref{fig:sdss_abs_2mass_mag}. The fit was based on stars with spectral types M0--L4.}
\end{deluxetable*}

We also examined the relationships between SDSS $riz$ and 2MASS $JHK_s$ absolute magnitudes as a function of the $(r-z)$ color. We selected $(r-z)$ as the base color because it is a good indicator of spectral type/effective temperature for M dwarfs. We divided the $(r-z)$ axis in intervals of $0.5$ mag and calculated the mean value and standard deviation for each interval. We fit the mean values with a third degree polynomial using the standard deviations as weights. We performed a fit for $0.5<r-z< 4.5$ mag (corresponding to M0--M9 dwarfs) because for L dwarfs, the main sequence turns over, as shown in Figure~\ref{fig:abs_mag_color_sdss}. This is the same break shown in the stellar color locus analysis in Figure \ref{fig:locus} at color $(G-\grp) \sim 1.7$. It means the relation between absolute magnitude and color cannot be used beyond this point because the two quantities are no longer related in the same way. The best fit parameters are in Table~\ref{table:fit_abs_mag_color_sdss}. 

We find that $M_{\rm r}$ has the tightest relation with $(r-z)$ color. The spread above and below the main sequence increases for the $i$ and $z$-bands. Furthermore, all three 2MASS bands (right panels) have higher spread above and below the main sequence than SDSS bands, and it also increases for the $K_s$ bands in comparison with $J$ and $H$. We compared our data and fit in $M_{\rm r}$ versus $(r-z)$ to the fit from \cite{Bochanski2010} as a check on our accuracy. The two fits are in good agreement, and both fall over the highest density of data points. 

\begin{deluxetable*}{cccccc}[ht!]
\tablecaption{Best fit parameters for SDSS and 2MASS magnitudes as a function of color. \label{table:fit_abs_mag_color_sdss}}
\tablehead{\colhead{Band} & \colhead{$a$} & \colhead{$b$} & \colhead{$c$} & \colhead{$d$} & \colhead{$\sigma$}}
\startdata
$M_{\rm r}$ & $-0.1 \pm 0.08$ & $0.63 \pm 0.68$ & $1.49 \pm 1.62$ & $6.35 \pm 1.1$ & $0.45$ \\ 
$M_{\rm i}$ & $-0.1 \pm 0.07$ & $0.62 \pm 0.59$ & $0.94 \pm 1.41$ & $6.18 \pm 0.96$ & $0.46$ \\ 
$M_{\rm z}$ & $-0.09 \pm 0.06$ & $0.59 \pm 0.5$ & $0.68 \pm 1.21$ & $6.1 \pm 0.83$ & $0.45$ \\ 
$M_{\rm J}$ & $-0.09 \pm 0.06$ & $0.58 \pm 0.45$ & $0.54 \pm 1.1$ & $5.08 \pm 0.76$ & $0.45$ \\ 
$M_{\rm H}$ & $-0.08 \pm 0.05$ & $0.49 \pm 0.44$ & $0.8 \pm 1.09$ & $4.26 \pm 0.76$ & $0.46$ \\ 
$M_{\rm Ks}$ & $-0.08 \pm 0.05$ & $0.49 \pm 0.42$ & $0.72 \pm 1.04$ & $4.15 \pm 0.72$ & $0.45$ \\ 
\enddata
\tablecomments{Results from the best fit to SDSS and 2MASS absolute magnitudes as function of color with a third degree polynomial, $M = a\times (r-z)^3 + b\times (r-z)^2 + c\times (r-z)+ d$, as shown in Figure~\ref{fig:sdss_abs_2mass_mag}. The fit is valid between $0.5<r-z<4.5$.}
\end{deluxetable*}

\pagebreak

\section{Age-related parameters}
\label{sec:age}

One of the primary goals for analysis of the MLSDSS-GaiaDR2 sample is to calibrate observable age indicators for M and L dwarfs. In this section, we examine the following activity-related and kinematic age indicators: (1) fractional $\halpha$ luminosity ($\lhalbol$), (2) vertical velocity dispersion ($\sigma _W$), (3) vertical action dispersion ($\sigma _{J_Z}$), and (4) tangential velocity ($v_t$). The relationship between $\halpha$, kinematics, and age has been explored in previous works \citep[e.g.,][]{West2008a,Pineda2013}. However our kinematics significantly improve the $20\%$ uncertainties on MLSDSS data as \textit{Gaia} DR2 contains proper motions with uncertainties of $1\%$ and distances with uncertainties of $5\%$.

\subsection{Fractional $\halpha$ luminosity on the Gaia color--magnitude diagram} 
\label{subsection:halphcmd}

\begin{figure*}[ht!]
\begin{center}
\includegraphics[width=\linewidth]{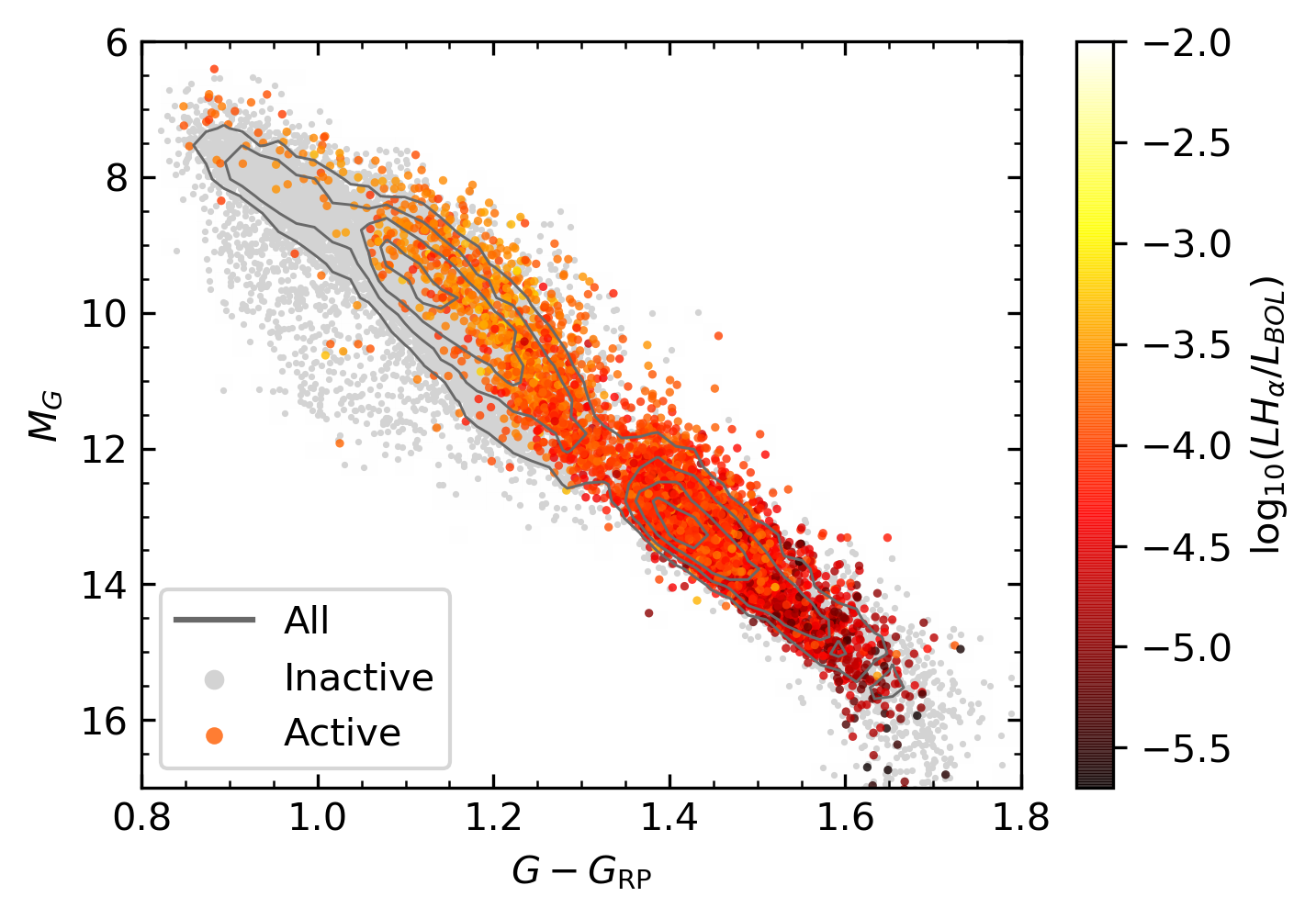}
\caption{\textit{Gaia} color--magnitude diagram, with magnetically active stars color coded by fractional $\halpha$ luminosity and inactive stars shown for comparison (grey points). We also include a contour plot for the full MLSDSS-GaiaDR2 sample. Extinction caused by dust between the star and the observer might inadvertently drive colors redward for non-young stars, therefore we removed stars which required a high extinction correction in SDSS photometry ($E(r-z)>0.1$).}
\label{fig:G_RP_Gabs_halpha}
\end{center}
\end{figure*}

Fractional $\halpha$ luminosity ($\lhalbol$) is a parameterization of the strength of the chromospheric $\halpha$ emission line which removes the dependence on the continuum that is a factor with EW measurements. This fractional $\halpha$ luminosity is an age indicator because it is a measure of stellar magnetic activity, which is presumed to be age-dependent: young stars have higher magnetic activity while old stars are less active or inactive \citep[e.g.,][]{Skumanich1972,Baliunas1995,Donahue1996,West2008, Mamajek2008}.
We show the relationship between fractional $\halpha$ luminosity and the position of the star on the \textit{Gaia} color--magnitude diagram in Figure~\ref{fig:G_RP_Gabs_halpha}, including inactive stars for comparison. To mitigate the effects of dust extinction which can also scatter stars on the color--magnitude diagram, we remove stars with high SDSS extinction, meaning $E(r-z)>0.1$.

As shown in Figure~\ref{fig:G_RP_Gabs_halpha}, the majority of the low mass, red stars ($G-\grp>1.3$; $>$M5), are both active and fall along the main sequence. The high fraction of active stars is due to the long ($\sim$7 Gyr) active lifetimes of late-M and L dwarfs \citep{Gizis2000,West2004,West2008,Schmidt2015}. 

\begin{figure}[ht!]
\begin{center}
\includegraphics[width=\linewidth]{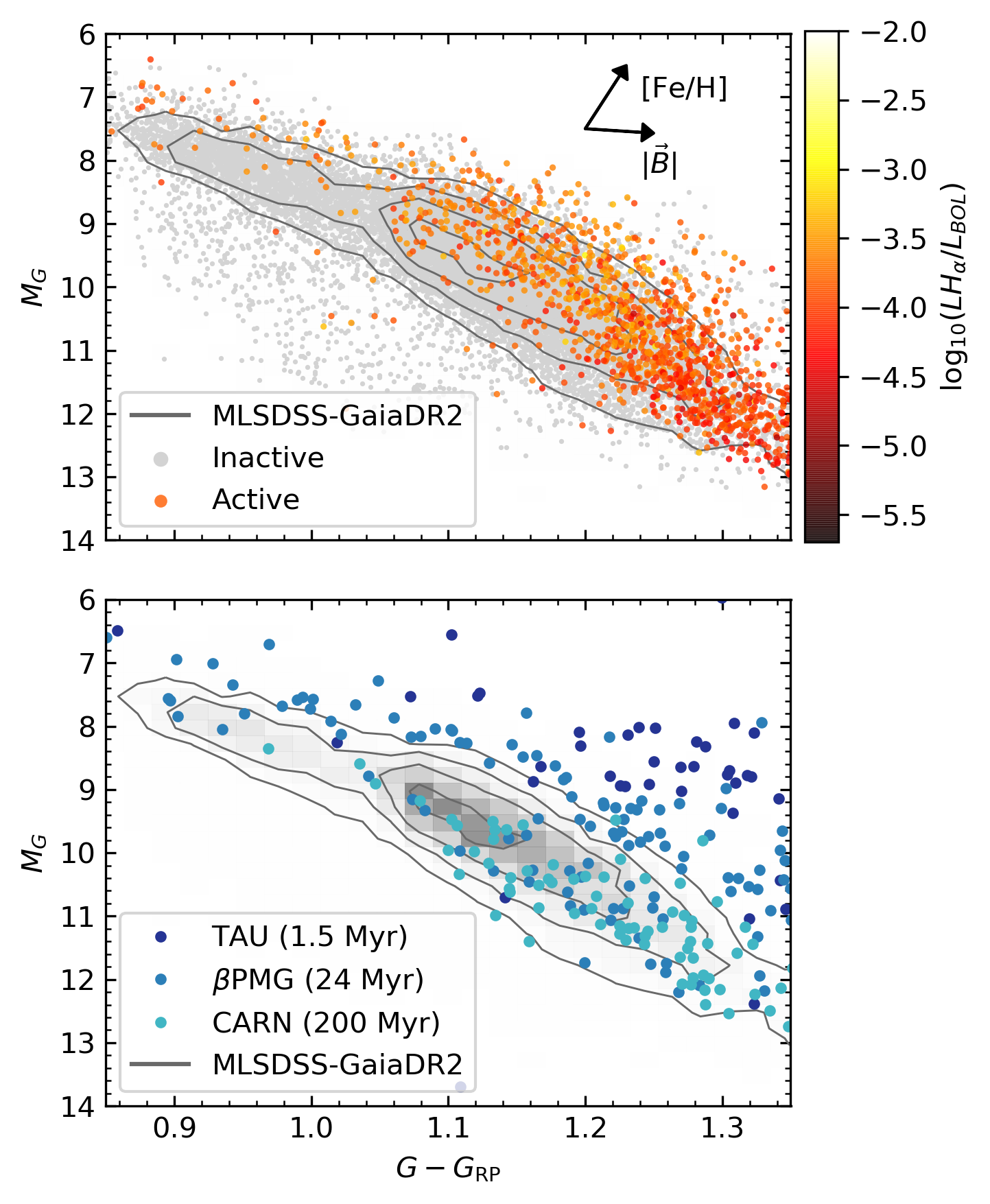}
\caption{Top panel: Zoom in of Figure~\ref{fig:G_RP_Gabs_halpha}. We included a reference to the effect of metallicity and radius inflation due to magnetic activity (lower effective temperature at approximately fixed luminosity) on the position of stars in the color--magnitude diagram with two arrows. These show an approximate direction for increasing metallicity ($Z$) and constant bolometric luminosity ($|\vec{B}|$) calculated from \citet{Mann2015}. Lower panel: We show the same contour plot for the full sample as in the top panel with three known moving groups members with known ages: Taurus (TAU, $1-2$ Myr), $\beta$ Pictoris ($\beta$PMG, $24\pm 3$ Myr) and Carina-Near (CARN, $\sim 200$ Myr). Active stars on the top panel look as young as $24$~Myr, but as this is not consistent with what we know from our sample we believe metallicity and, for the active stars, radius inflation are the causes of the scattered data.}
\label{fig:G_RP_Gabs_halpha_zoom_mg}
\end{center}
\end{figure}

We find a clear correlation between activity and the position in the color--magnitude diagram for the bluer ($(G-\grp)<1.3$;$<$M5) stars and in Figure~\ref{fig:G_RP_Gabs_halpha_zoom_mg}, we zoom into this region.
Active stars are found, on average, at redder colors and/or brighter magnitudes than inactive stars. 
The four most probable causes are youth, metallicity, binarity and/or magnetic activity. The effects of each are described below. 

To investigate the effect of youth, we compared the color--magnitude position of the active stars with the position of three known young moving groups from \citet{Gagne2018a}: Taurus (TAU, $1-2$~Myr), $\beta$~Pictoris ($\beta$PMG, $24\pm3$~Myr) and Carina-Near (CARN, $\sim200$~Myr; Figure~\ref{fig:G_RP_Gabs_halpha_zoom_mg}). 
The oldest of the moving groups, CARN, is the closest to the main sequence, while the younger groups fall above it, mostly due to the stars still contracting and having larger radii than stars of the same mass that have reached the main sequence. 
While these young stars are active, they are on average not as active as the most strongly active stars in the MLSDSS-GaiaDR2 sample. Therefore we speculate their position above the main sequence is primarily due to their pre-main sequence radius rather than their activity level. 
The comparison between these young stars and the active MLSDSS-GaiaDR2 stars provides an estimate of how much radius inflation due to youth is responsible for their position on the color--magnitude diagram. 

In Figure~\ref{fig:G_RP_Gabs_halpha_zoom_mg}, we show that active MLSDSS-GaiaDR2 stars that lie just above the main sequence have approximately the same position as the $24$~Myr moving group $\beta$PMG. 
However, these active stars are unlikely to be young: 
they are within $200$~pc from the Sun 
but there are a limited number of associations at or around $24$~Myr at these nearby distances (e.g. $32$~Orionis, see \citealt{Faherty2018}) and none of these stars appear to be members of known young groups. 
Furthermore, most of the early M dwarfs in MLSDSS-GaiaDR2 are highly separated from the plane of the Galaxy which indicates they are old \citep[e.g.,][]{West2004}. 
Therefore the position of the active stars at redder colors and/or brighter magnitudes than inactive stars is not due entirely to youth. 

Binarity could also affect the position of active stars in the color--magnitude diagram. Tight binaries are more luminous and could be more active due to tidal interactions \citep{Shkolnik2011}.

Another factor that influences the position of stars on the color-magnitude diagram is its metallicity. \citet{Mann2015} showed that high metallicity M dwarfs tend to have larger radii than low metallicity M~dwarfs for a given effective temperature. This effect could be another factor in the position of active stars on the color-magnitude diagram.

Moreover, \citet{Bochanski2011} showed that active stars fall even redder and/or higher above the main sequence than inactive stars with the same metallicity. Active stars have been shown to have inflated radii, possibly caused by strong surface magnetic fields \citep[e.g.][]{Lopez-Morales2005,Morales2009,Torres2013}.
Observations of young low-mass stars show radii that are, at a fixed mass, $10$--$15\%$ larger than predicted by evolutionary models \citep[e.g.][]{Somers2017,Cruz2018,Kesseli2018}.
\citet{Stassun2012} showed that for low mass stars, the activity strength of active stars (as indicated by $\lhalbol$ and $\lxraylbol$) is correlated with inflated radii and cooler effective temperatures compared to inactive stars.
They also found that radius inflation and cooler temperatures cancel the effect of magnetic activity on bolometric luminosity, meaning that these effects should mostly cause horizontal shifts on the color magnitude diagram, not vertical ones.

Youth, binarity, metallicity, and activity can all play a role in scattering M0--M5~stars, shown in the top panel of Figure~\ref{fig:G_RP_Gabs_halpha_zoom_mg} to redder colors and/or brighter magnitudes than the bulk of the main sequence. Metallicity and binarity effects seem combined with activity to lift stars further off the main sequence, however, it is unlikely our active stars are particularly young. Magnetism likely plays a strong role in the position of active stars on the color-magnitude diagram. 
This effect might also exist for the later spectral types ($>$M5) but it is not evident in our current analysis.


\subsection{Vertical velocity and vertical action dispersion}
\label{subsec:verticalvelocity}

Full three-dimensional space motion has been shown to trace stellar ages in the Galaxy. As stars age, increased interactions with giant molecular clouds and passing stars result in kinematic heating.  Therefore one can use the overall velocity distribution of a population of stars to infer the age of that population via an age-velocity dispersion relation \citep[AVR; e.g.][]{Wielen1977,Hanninen2002}.
Previous works have used full kinematics, or tangential velocity as a proxy for full space motion, to estimate the kinematic age of the low-mass star population compared to higher mass stars \citep[e.g.][]{Schmidt2007,ZapateroOsorio2007,Faherty2009,Reiners2009}.
Vertical action is related to the vertical component of a star's angular momentum integrated over the gravitational potential of the Milky Way. Previous work has showed that in particular, vertical velocity and vertical action dispersion ($\sigma _W$ and $\sigma _{J_Z}$) are correlated with age \citep[e.g.][]{West2006,West2008,Nordstrom2004,Aumer2016,Yu2018}. As we do not have ages for the stars in the MLSDSS-GaiaDR2 sample yet, we studied the correlation between $\sigma _W$ and $\sigma _{J_Z}$ and fractional $\halpha$ luminosity, another age indicator (see Section \ref{subsection:halphcmd} for the $\halpha$ analysis).

We calculated vertical actions and vertical velocities using positions, proper motions and parallaxes from \textit{Gaia} DR2 and radial velocities from MLSDSS. Note that the stars in the MLSDSS-GaiaDR2 sample are too faint to have radial velocities in \textit{Gaia} DR2. For this analysis we used the good astrometric sample described in Section \ref{subsubsec:astrometriccuts} and we added cuts for radial velocity signal to noise: $\texttt{rv/rv\_err} > 2$ and absolute value: $\| \texttt{rv}\| < 500$ km/s. The number of stars after the extra cuts for radial velocity is $15,988$ ($67\%$ of the good astrometric sample of $23,842$ stars). We also removed stars categorized as white dwarf-M dwarf binaries because the white dwarf can affect the magnetic field of the companion (see Section \ref{subsec:basemlsdss}, \citealt{Morgan2012}). To compute vertical velocities and vertical actions we used Galpy\footnote{\url{http://github.com/jobovy/galpy}} \citep{Binney2012,Bovy2013,Bovy2015} and W. Trick's code\footnote{\url{https://github.com/wilmatrick/GaiaSprint/blob/master/Action_Galpy_Tutorial.ipynb}} with the Milky Way potential from \citet{Bovy2015}.
Uncertainties on these values were computed via Monte Carlo.

To compute the dispersion, we divided the values of logarithmic fractional $H\alpha$ luminosity
($\log_{10}\left(\lhalbol \right)$) into six 
regularly spaced bins and calculated the dispersion per bin ($\sigma _{bin}$).
The value of fractional $\halpha$ luminosity assigned to each $\sigma _{bin}$ corresponds 
to the middle of the bin. 
To calculate the dispersion per bin we used the median absolute deviation to alleviate the influence of outliers.
Uncertainties on the median absolute deviation were estimated, again by performing Monte Carlo
re-sampling of data points within their uncertainties.



\begin{figure}[ht!]
\begin{center}
\includegraphics[width=\linewidth]{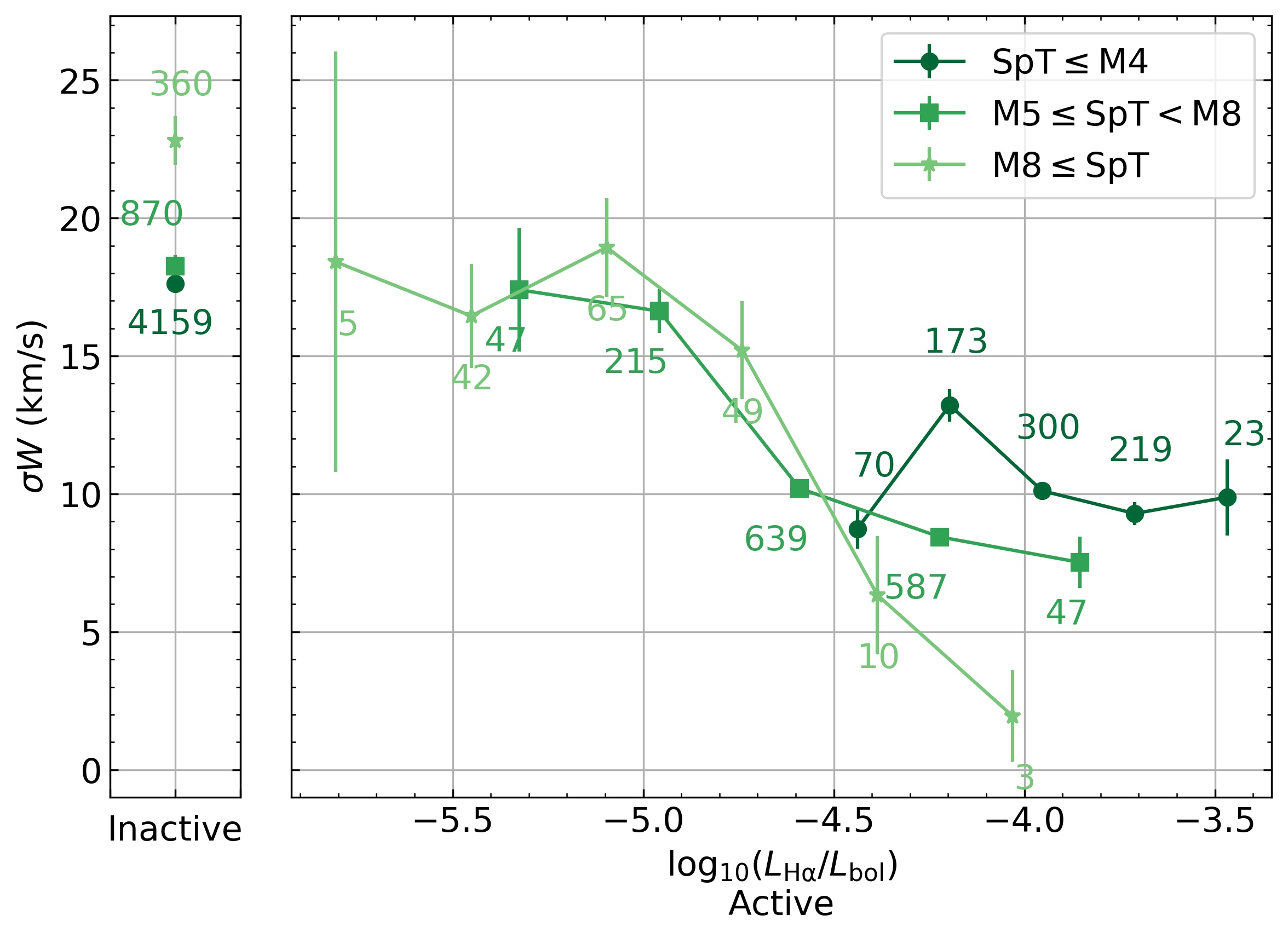}
\caption{Vertical velocity dispersion ($\sigma _W$) as a function of fractional $\halpha$ luminosity.
The dispersion is the median absolute deviation of points within each bin.
The left panel contains the dispersion calculated for all the stars classified as 
inactive in MLSDSS and the right panel the active ones. The number of stars that 
were used to calculate the dispersion value is indicated next to each point.
We used the Good Astrometry sample described in Section \ref{subsubsec:astrometriccuts}
plus cuts for the radial velocity (see~\ref{subsec:verticalvelocity}).}
\label{fig:dispersion_vv}
\end{center}
\end{figure}

Results for the dispersion of vertical velocity as a function of $\halpha$ luminosity are presented in Figure~\ref{fig:dispersion_vv}. We divided the data into three spectral type bins: SpT$\leq$M4, M5$\leq$SpT$<$M8 and M8$\leq$SpT, as well as two categories of active and inactive stars (see Section \ref{subsec:basemlsdss} for more detail on the classification of active and inactive). For the active stars, we find that $\sigma _W$ is lower for high $\halpha$ activity stars than for less active stars, and inactive stars have a higher vertical velocity dispersion than active stars on average.
Magnetically active stars are younger than less active or inactive stars
\citep[e.g.,][]{Skumanich1972,West2008}, therefore Figure~\ref{fig:dispersion_vv} 
is showing that vertical velocity is also correlated with age: young stars have a 
smaller vertical velocity dispersion because they have had less time to experience
orbital perturbations in the $Z$ direction (out of the galactic plane).

The activity-velocity dispersion relation does not show an obvious dependence on spectral type which is a proxy for mass 
for mid and late dwarfs (SpT $>M5$).
Active early-M dwarfs have higher vertical velocity dispersions compared to later type dwarfs. This is likely due to the detection threshold for the proper motions of the most distant M dwarfs; those with lower tangential velocities have lower quality proper motions, so only stars with high velocities have reliable proper motions, therefore biasing the dispersion to larger values \citep{Bochanski2011}. Moreover, the sample of early-M dwarfs is biased towards old stars compared to the sample of mid- to late-M dwarfs due to selection effects. The SDSS photometric detectors saturated for sources brigher than $14$ mag in $r$, they cannot obtain reliable photometry for $M0-M4$ dwarfs found closer than $100-200$~pc to the Sun. On the faint end, SDSS spectra only have sufficient quality to be included in the sample of objects brighter than $\sim23$~mag in $r$, including early-M dwarfs as distant as $\sim1-2$~kpc but late-M dwarfs only are detected at a distance of $\sim 100-200$~pc. As a consequence, early-M dwarfs found in the MLSDSS-GaiaDR2 sample are typically higher above the plane of the Galaxy and so they are likely older than later-type M dwarfs.



\begin{figure}[ht!]
\begin{center}
\includegraphics[width=\linewidth]{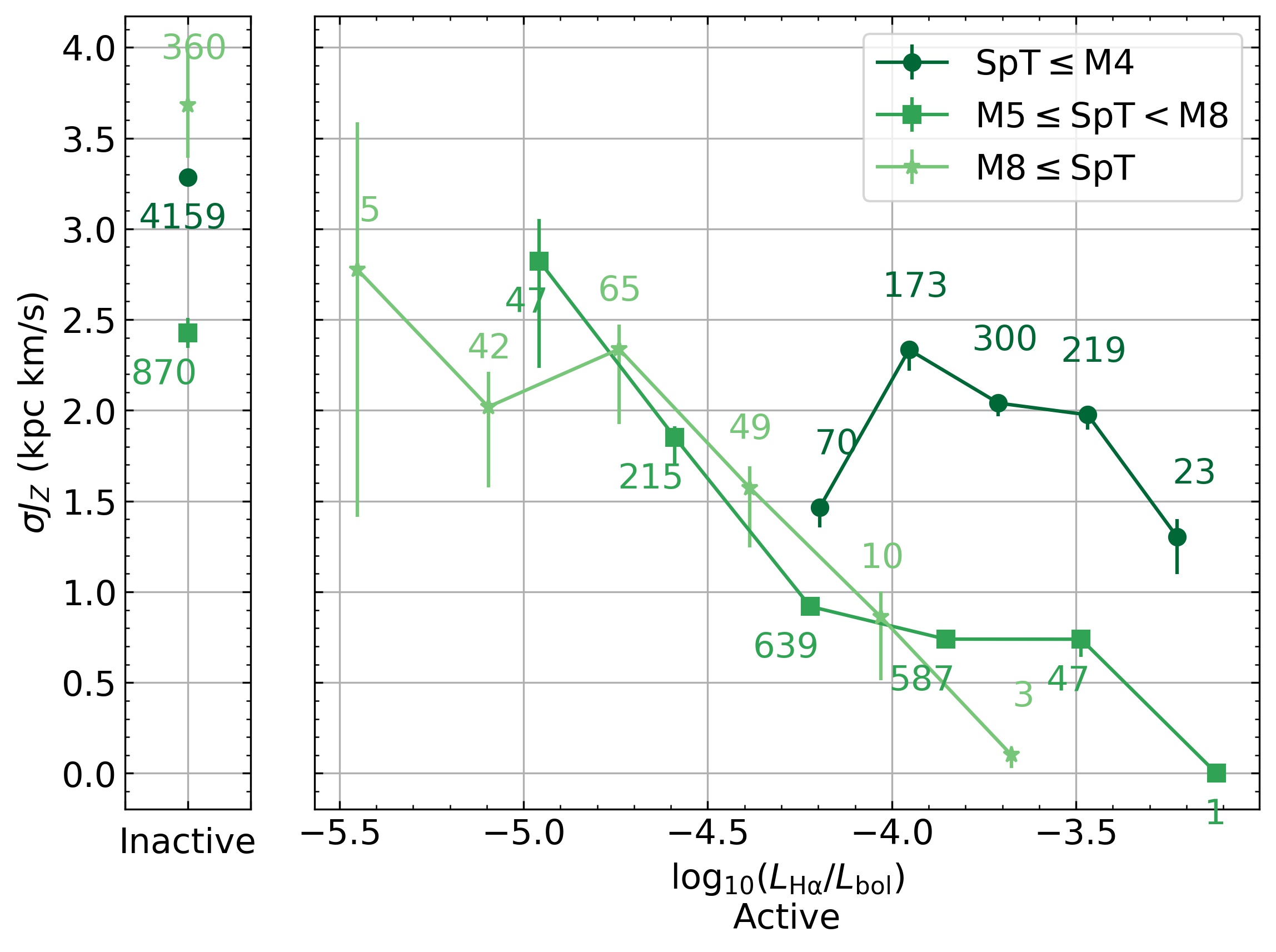}
\caption{Vertical action dispersion ($\sigma _{J_Z}$) as a function of fractional $\halpha$ luminosity. See caption of Figure \ref{fig:dispersion_vv}. The error-bars are the $16$ and $64$ percentiles.}
\label{fig:dispersion_va}
\end{center}
\end{figure}

We calculated vertical action using a similar procedure used to calculate vertical
velocity, and Figure~\ref{fig:dispersion_va} shows vertical action dispersion as a
function of fractional $\halpha$ luminosity. In this case, not all of the vertical action distributions are gaussian, so the distributions of dispersion per bin are also not a gaussian. Therefore, we represent the uncertainties with the $16$ and $64$ percentiles.
Similarly to the vertical velocity analysis, 
inactive stars have significantly larger vertical action dispersion than active
stars, and for active stars, the dispersion decreases with increasing $\halpha$ activity.
This indicates that vertical action
dispersion, similar to the vertical velocity dispersion, is another age indicator: young stars have low vertical action dispersion
while old stars have higher dispersion because they were kinematically heated.
Early M-dwarfs in Figure~\ref{fig:dispersion_va} seem to have higher vertical {\it action} dispersion; however, this is likely due to the same biases from distant M~dwarfs explained above.

Figures~\ref{fig:dispersion_vv} and \ref{fig:dispersion_va} are the first steps to
obtaining a functional description of how kinematics indicate the age of low mass stars
(e.g. work such as \citet{Wielen1977} for higher mass stars).


\subsection{Tangential velocity} 
\label{subsec:tanvel}

\begin{figure}[ht!]
\begin{center}
\includegraphics[width=\linewidth]{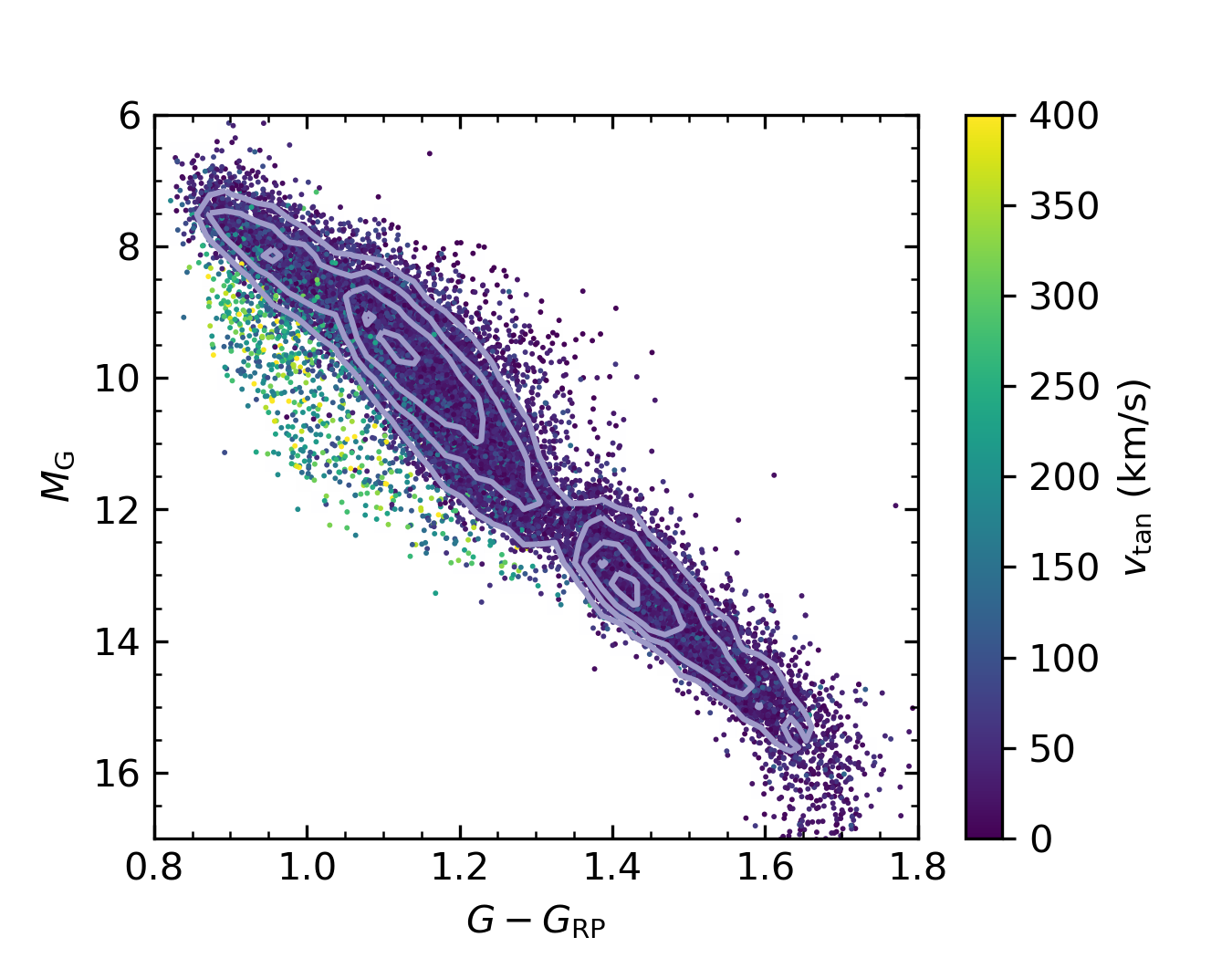}
\caption{Color magnitude diagram color coded with tangential velocity. Stars below the main sequence have high tangential velocity and are inactive stars as shown in Figure \ref{fig:G_RP_Gabs_halpha}. This is consistent with being low metallicity stars and halo or thick disk stars. These are indicators that they are old stars.}
\label{fig:spt_Gabs_vt}
\end{center}
\end{figure}

With \textit{Gaia} DR2 we were able to calculate precise tangential velocities  
for $22,373$ M and L dwarfs in the MLSDSS-GaiaDR2 sample.
To explore the disk and halo populations of stars in our catalog, we studied the correlation between tangential velocity ($v_{\rm tan}$) and 
color--magnitude diagram position for the MLSDSS-GaiaDR2 sample (Figure \ref{fig:spt_Gabs_vt}) as done by previous work \citep{Lepine2007,Gizis1999,Jao2017}.

There is a significant number of stars below the main sequence for $(G-\grp)<1.3$ 
which have a high tangential velocity, with $\overline{v}_{\rm tan} \sim 200$ km/s, in 
comparison to the rest of the stars in the sample, with $\overline{v}_{\rm tan} \sim 50$ km/s.
Such objects that are blue and fast are likely members of the older thick disk or halo.
At least half of these stars were classified as subdwarf candidates with the cuts in
\citet{Gizis1999} and \citet{Jao2017}.
Furthermore, our stars are in the same place in the color--magnitude diagram as the
subdwarfs in \citet{Jao2017}, and $376$ of them were classified as subdwarfs by \citet{Savcheva2014} (See Section \ref{subsec:cmdmlsdssgaiadr2}).
These high tangential velocity objects also have low fractional $\halpha$
luminosities (see Figure \ref{fig:G_RP_Gabs_halpha}).
The lack of magnetic activity paired with high tangential velocities, low metallicities, and
blue optical colors affirms they are likely an older population of low mass stars.

\section{Summary}
\label{sec:summary}
We compiled the MLSDSS-GaiaDR2 sample of $74,216$ M and L dwarfs. To create the sample we combined two SDSS catalogs: the SDSS DR7 M dwarfs spectroscopic catalog \citep{West2011} and the BUD catalog \citep[][in prep]{Schmidt2015,schmidt2019}, into the MLSDSS sample. $\halpha$ equivalent widths, spectral types and radial velocities were calculated by the authors of these catalogs. We cross-matched the MLSDSS sample with \textit{Gaia} DR2 to obtain proper motions and parallaxes for the stars. We found $73,003$ matches and we used a very conservative criterion to identify $67,573$ good matches. We adjusted some of the quality cuts suggested by the \textit{Gaia} Papers \citep[e.g.][]{Lindegren2018,Evans2018,Arenou2018,Collaboration2018a} to make them suitable for the later M and L dwarfs.  The final MLSDSS-GaiaDR2 sample contains $\halpha$ equivalent widths, spectral types, SDSS, 2MASS and \textit{Gaia} photometry and proper motions and parallaxes from \textit{Gaia} DR2. The good astrometric sample contains $23,842$ stars and the good photometry sample for the $G$, $\grp$ and $\gbp$ bans have $22,706$, $22,373$ and $16,527$ stars respectively.

With the MLSDSS-GaiaDR2 sample we calculated mean absolute magnitudes and colors as a function of spectral type using the three photometric bands in \textit{Gaia} DR2: $G$, $\grp$ and $\gbp$. Furthermore, we characterized the color-color space of the stellar locus for \textit{Gaia}, SDSS and 2MASS colors. We used the distances calculated with \textit{Gaia} DR2 parallaxes, which are one order of magnitude better than the photometric distances from MLSDSS, to plot the color--magnitude diagram for the MLSDSS-GaiaDR2 sample. We found that the main sequence widens as it goes towards cooler stars. This effect starts around spectral type M3 
and could be related to the transition to fully convective interior. We also used the the MLSDSS-GaiaDR2 SDSS and 2MASS photometry to calculate absolute magnitudes as a function of spectral type and color for the $riz$-photometry and $J$, $H$ and $K_s$ bands. We compared our results with \citet{Hawley2002} and \citet{Bochanski2011} and found good agreement.

We noticed that active stars are found, on average, at redder colors and/or brighter magnitudes than inactive stars in the color--magnitude diagram. Comparing to the position in the color--magnitude diagram of three young moving groups with different ages we found that youth alone cannot explain the position of active stars. The stars in the MLSDSS-GaiaDR2 sample are mostly high above the galactic plane, therefore unlikely to be young. We hypothesize that the position of active stars might be due to binarity, metallicity and/or that magnetism likely plays a strong role by inflating the radii of the stars and reducing their effective temperature, which makes them look redder. 

Furthermore, we found that early types of inactive stars that are faint in absolute magnitude (below the main sequence) have high tangential velocities ($\sim 150$ km/s), which indicates they belong to the halo or thick disk and that they are an old population of M dwarfs. Furthermore, $376$ of these were identified as subdwarfs by \citet{Savcheva2014}.  

Finally, we studied the relation between vertical velocity and vertical action dispersion with fractional $\halpha$ luminosity. We found that stars with higher $\halpha$ activity have lower dispersion both in vertical velocity and vertical action and stars with lower $\halpha$ activity or inactive have higher dispersion. As $\halpha$ is an age indicator, this means that young (active) stars live close to the plane of the galaxy, so their vertical action and vertical velocity dispersion is small, and old (less active or inactive) stars were kinematically heated, so their dispersion is higher. In future work we will fit these relations using bayesian inference and we will constrain the ages of M and L dwarfs using the age indicators in the MLSDSS-GaiaDR2 sample. 

\section{Acknowledgements}

The authors would like to thank Andrew Mann, John Bochanski, Ricky Smart, Jonathan Gagn\'e and Eric Mamajek for useful discussions and helpful comments on this work. The authors would also like to thank the referee for their comments, which were very useful in clarifying and improving the analysis and results.

This work is supported by: SDSS Faculty and Student Team (FAST) initiative, NASA through the American Astronomical Society's Small Research Grant Program, the American Museum of Natural History (AMNH) and the National Science Foundation under Grant No. 1614527.
This publication makes use of data products from the Two Micron All Sky Survey, which is a joint project of the University of Massachusetts and the Infrared Processing and Analysis Center/California Institute of Technology, funded by the National Aeronautics and Space Administration and the National Science Foundation. 
Funding for SDSS-III has been provided by the Alfred P. Sloan Foundation, the Participating Institutions, the National Science Foundation, and the U.S. Department of Energy Office of Science. The SDSS-III web site is http://www.sdss3.org/. SDSS-III is managed by the Astrophysical Research Consortium for the Participating Institutions of the SDSS-III Collaboration including the University of Arizona, the Brazilian Participation Group, Brookhaven National Laboratory, University of Cambridge, Carnegie Mellon University, University of Florida, the French Participation Group, the German Participation Group, Harvard University, the Instituto de Astrofisica de Canarias, the Michigan State/Notre Dame/JINA Participation Group, Johns Hopkins University, Lawrence Berkeley National Laboratory, Max Planck Institute for Astrophysics, Max Planck Institute for Extraterrestrial Physics, New Mexico State University, New York University, Ohio State University, Pennsylvania State University, University of Portsmouth, Princeton University, the Spanish Participation Group, University of Tokyo, University of Utah, Vanderbilt University, University of Virginia, University of Washington, and Yale University.  Funding for the SDSS and SDSS-II has been provided by the Alfred P. Sloan Foundation, the Participating Institutions, the National Science Foundation, the U.S. Department of Energy, the National Aeronautics and Space Administration, the Japanese Monbukagakusho, the Max Planck Society, and the Higher Education Funding Council for England. The SDSS Web Site is http://www.sdss.org/. The SDSS is managed by the Astrophysical Research Consortium for the Participating Institutions. The Participating Institutions are the American Museum of Natural History, Astrophysical Institute Potsdam, University of Basel, University of Cambridge, Case Western Reserve University, University of Chicago, Drexel University, Fermilab, the Institute for Advanced Study, the Japan Participation Group, Johns Hopkins University, the Joint Institute for Nuclear Astrophysics, the Kavli Institute for Particle Astrophysics and Cosmology, the Korean Scientist Group, the Chinese Academy of Sciences (LAMOST), Los Alamos National Laboratory, the Max-Planck-Institute for Astronomy (MPIA), the Max-Planck-Institute for Astrophysics (MPA), New Mexico State University, Ohio State University, University of Pittsburgh, University of Portsmouth, Princeton University, the United States Naval Observatory, and the University of Washington.  
 
This work has made use of data from the European Space Agency (ESA) mission {\it Gaia} (\url{https://www.cosmos.esa.int/gaia}), processed by the {\it Gaia} Data Processing and Analysis Consortium (DPAC, \url{https://www.cosmos.esa.int/web/gaia/dpac/consortium}). Funding for the DPAC has been provided by national institutions, in particular the institutions participating in the {\it Gaia} Multilateral Agreement.

Part of this research was done at the NYC \textit{Gaia} DR2 Workshop at the Center for Computational Astrophysics of the Flatiron Institute in 2018 April.

\software{Astropy~\citep{astropy:2013, astropy:2018},~Galpy~\citep{Binney2012,Bovy2013,Bovy2015},~Matplotlib~\citep{Hunter:2007},~NumPy~\citep{van2011numpy},~SciPy~\citep{jones_scipy_2001}~and~TOPCAT \citep{2005ASPC..347...29T}.} 

\bibliographystyle{aasjournal}
\bibliography{references.bib,extras.bib}

\end{document}